\documentclass[acmtog,authorversion]{acmart}
\usepackage{geometrycollective}
\usepackage{mdframed}

\hypersetup{
    colorlinks = false, % set to true for editing symbols and links
    linkcolor = blue,
    citecolor = black
}

\DeclareMathAlphabet{\mathcal}{OMS}{cmsy}{m}{n}

\usepackage{wrapfig}
\usepackage{picinpar}

\usepackage{enumerate}

\usepackage{booktabs} % For formal tables

\usepackage{algorithm}
\usepackage{algorithmicx}
\usepackage{algpseudocode}

\algtext*{EndProcedure}
\algtext*{EndIf}
\algtext*{EndFor}

\renewcommand{\algref}[1]{Alg.~\ref{alg:#1}}

\newcommand{\InputConditions}[1]{\Require \parbox[t]{\dimexpr\linewidth-\algorithmicindent}{#1\strut}}
\newcommand{\OutputConditions}[1]{\Ensure \parbox[t]{\dimexpr\linewidth-\algorithmicindent}{#1\strut}}

\usepackage{tikz}
\DeclareRobustCommand{\circled}[1]{%
   \tikz[baseline=(char.base)]{
      \node[shape=circle,draw,semithick,inner sep=.5pt] (char) {#1};}
   }

\acmJournal{TOG}
\acmVolume{37}
\acmNumber{1}
\acmArticle{5}
\acmYear{2017}
\acmMonth{12}

\setcopyright{usgovmixed}
\acmDOI{0000001.0000001_2}

\newcommand{\tA}{\tilde{A}}
\newcommand{\dtA}{d\tA}

\newcommand{\Qavg}{\mathcal{Q}_{\mathrm{avg}}}
\newcommand{\Qc}{\mathcal{Q}}
\newcommand{\on}{\mathrm{on}\quad}

\newcommand{\II}{{\mathsf{I}\mathsf{I}}}
\newcommand{\IB}{{\mathsf{I}\mathsf{B}}}
\newcommand{\BI}{{\mathsf{B}\mathsf{I}}}
\newcommand{\BB}{{\mathsf{B}\mathsf{B}}}

\usepackage[refpage]{nomencl}

\makenomenclature

\makeatletter
 \newcommand{\hyperraisedtarget}[1]{\Hy@raisedlink{\hypertarget{#1}{}}}
\makeatother

\newcommand{\definesymbol}[3]{\nomenclature{\(#1\)}{#2}\hyperraisedtarget{symb:#3}}

\newcommand{\M}                     {{\protect\hyperlink{symb:surface}{M}}}
\newcommand{\bM}                    {{\protect\hyperlink{symb:boundary}{\partial M}}}
\newcommand{\J}                     {{\protect\hyperlink{symb:complexstructure}{J}}}
\newcommand{\f}                     {{\protect\hyperlink{symb:flattening}{f}}}
\newcommand{\df}                    {{\protect\hyperlink{symb:differential}{d\!f}}}
\newcommand{\n}                     {{\protect\hyperlink{symb:normal}{n}}}
\newcommand{\T}                     {{\protect\hyperlink{symb:tangent}{T}}}
\newcommand{\K}                     {{\protect\hyperlink{symb:gausscurvature}{K}}}
\newcommand{\innervec}[2]           {{\protect\hyperlink{symb:innervec}{\langle #1, #2 \rangle}}}

\let\oldimath\imath
\renewcommand{\imath}               {{\protect\hyperlink{symb:imaginaryunit}{\oldimath}}}
\let\oldarg\arg
\renewcommand{\arg}                 {{\protect\hyperlink{symb:argument}{\oldarg}}}
\newcommand{\tipangle}              {{\protect\hyperlink{symb:tipangle}{\beta}}}
\let\oldkappa\kappa
\renewcommand{\kappa}               {{\protect\hyperlink{symb:geodesiccurvature}{\oldkappa}}}
\newcommand{\tksf}                  {{\protect\hyperlink{symb:newexteriorangles}{\tilde{\mathsf{k}}}}}
\newcommand{\tkappa}                {{\protect\hyperlink{symb:newgeodesiccurvature}{\tilde{\oldkappa}}}}
\newcommand{\mesh}                  {{\protect\hyperlink{symb:mesh}{\Msf}}}
\newcommand{\vertices}              {{\protect\hyperlink{symb:vertices}{\Vsf}}}
\newcommand{\bvertices}             {{\protect\hyperlink{symb:boundaryvertices}{\Bsf}}}
\newcommand{\ivertices}             {{\protect\hyperlink{symb:interiorvertices}{\Isf}}}
\newcommand{\edges}                 {{\protect\hyperlink{symb:edges}{\Esf}}}
\newcommand{\bedges}                {{\protect\hyperlink{symb:boundaryedges}{\partial\mathsf{M}}}}
\newcommand{\faces}                 {{\protect\hyperlink{symb:faces}{\Fsf}}}
\newcommand{\dualcell}              {{\protect\hyperlink{symb:dualcell}{C}}}
\let\oldgamma\gamma
\renewcommand{\gamma}               {{\protect\hyperlink{symb:curve}{\oldgamma}}}
\newcommand{\tgamma}                {{\protect\hyperlink{symb:newcurve}{\tilde{\oldgamma}}}}
\newcommand{\s}                     {{\protect\hyperlink{symb:arclength}{s}}}
\newcommand{\ts}                    {{\protect\hyperlink{symb:newarclength}{\tilde{s}}}}
\newcommand{\ds}                    {{\protect\hyperlink{symb:lengthdensity}{ds}}}
\newcommand{\tds}                   {{\protect\hyperlink{symb:newlengthdensity}{d\tilde{s}}}}
\newcommand{\dA}                    {{\protect\hyperlink{symb:areadensity}{dA}}}
\newcommand{\Htransform}            {{\protect\hyperlink{symb:hilberttransform}{\mathcal{H}}}}
\renewcommand{\u}                   {{\protect\hyperlink{symb:logfactor}{u}}}
\renewcommand{\a}                   {{\protect\hyperlink{symb:realpartf}{a}}}
\renewcommand{\b}                   {{\protect\hyperlink{symb:imaginarypartf}{b}}}
\renewcommand{\r}                   {{\protect\hyperlink{symb:adjustfactor}{r}}}
\newcommand{\g}                     {{\protect\hyperlink{symb:dirichletdata}{g}}}
\newcommand{\h}                     {{\protect\hyperlink{symb:neumanndata}{h}}}
\let\oldphi\phi
\renewcommand{\phi}                 {{\protect\hyperlink{symb:poissonrhs}{\oldphi}}}
\newcommand{\dbedge}                {{\protect\hyperlink{symb:dualboundaryedge}{e}}}
\newcommand{\conformalenergy}       {{\protect\hyperlink{symb:conformalenergy}{E_C}}}
\newcommand{\cotLaplace}            {{\protect\hyperlink{symb:cotanlaplace}{\Asf}}}
\renewcommand{\Lsf}                 {{\protect\hyperlink{symb:choleskyfactor}{\mathsf{L}}}}
\let\oldOmega\Omega
\renewcommand{\Omega}               {{\protect\hyperlink{symb:discretegausscurvature}{\oldOmega}}}
\renewcommand{\ksf}                 {{\protect\hyperlink{symb:discretegeodesiccurvature}{\mathsf{k}}}}
\let\oldDelta\Delta
\renewcommand{\Delta}               {{\protect\hyperlink{symb:laplacebeltrami}{\oldDelta}}}
\let\oldLambda\Lambda
\renewcommand{\Lambda}              {{\protect\hyperlink{symb:dirichlettoneumann}{\oldLambda}}}
\newcommand{\edgelength}[1]         {{\protect\hyperlink{symb:edgelength}{\ell_{#1}}}}
\newcommand{\duallength}[1]         {{\protect\hyperlink{symb:duallength}{\ell_{#1}}}}
\newcommand{\targetlength}[1]       {{\protect\hyperlink{symb:targetlength}{\ell^*_{#1}}}}
\newcommand{\finallength}[1]        {{\protect\hyperlink{symb:finallength}{\tilde{\ell}_{#1}}}}
\let\oldTheta\Theta
\renewcommand{\Theta}               {{\protect\hyperlink{symb:targetconeangles}{\oldTheta}}}
\renewcommand{\usf}                 {{\protect\hyperlink{symb:discretescalefactor}{\mathsf{u}}}}
\renewcommand{\hsf}                 {{\protect\hyperlink{symb:discreteneumanndata}{\mathsf{h}}}}
\renewcommand{\gsf}                 {{\protect\hyperlink{symb:discretedirichletdata}{\mathsf{g}}}}
\renewcommand{\Tsf}                 {{\protect\hyperlink{symb:discretetangent}{\mathsf{T}}}}
\newcommand{\tTsf}                  {{\protect\hyperlink{symb:targettangent}{\widetilde{\mathsf{T}}}}}
\renewcommand{\fsf}                 {{\protect\hyperlink{symb:discretemap}{\mathsf{f}}}}

\newcommand{\defineSurface}                   {\definesymbol{\M}                                {domain to be flattened (topological disk)}              {surface}}
\newcommand{\defineBoundary}                  {\definesymbol{\bM}                               {domain boundary}                                        {boundary}}
\newcommand{\defineComplexStructure}          {\definesymbol{\J}                                {90-degree rotation}                                     {complexstructure}}
\newcommand{\defineFlattening}                {\definesymbol{\f: \M \to \CC}                    {flattening}                                             {flattening}}
\newcommand{\defineDifferential}              {\definesymbol{\df}                               {differential of flattening \(\f\)}                      {differential}}
\newcommand{\defineNormal}                    {\definesymbol{\n}                                {normal to \(\bM\)}                                      {normal}}
\newcommand{\defineTangent}                   {\definesymbol{\T}                                {unit tangent along \(\bM\)}                             {tangent}}
\newcommand{\defineGaussCurvature}            {\definesymbol{\K}                                {Gaussian curvature of domain}                           {gausscurvature}}
\newcommand{\defineInnerVec}                  {\definesymbol{\innervec{\cdot}{\cdot}}           {real inner product of vectors}                          {innervec}}
\newcommand{\defineImaginaryUnit}             {\definesymbol{\imath}                            {imaginary unit (\(\imath^2 = -1\))}                     {imaginaryunit}}
\newcommand{\defineArgument}                  {\definesymbol{\arg}                              {complex argument (angle with real axis)}                {argument}}
\newcommand{\defineTipAngle}                  {\definesymbol{\tipangle_i^{jk}}                  {interior angle at corner \(i\) of triangle \(\ijk\)}    {tipangle}}
\newcommand{\defineMesh}                      {\definesymbol{\mesh}                             {input mesh}                                             {mesh}}
\newcommand{\defineVertices}                  {\definesymbol{\vertices}                         {vertices of \(\mesh\)}                                  {vertices}}
\newcommand{\defineBoundaryVertices}          {\definesymbol{\bvertices}                        {boundary vertices of \(\mesh\)}                         {boundaryvertices}}
\newcommand{\defineInteriorVertices}          {\definesymbol{\ivertices}                        {interior vertices of \(\mesh\)}                         {interiorvertices}}
\newcommand{\defineEdges}                     {\definesymbol{\edges}                            {edges of \(\mesh\)}                                     {edges}}
\newcommand{\defineBoundaryEdges}             {\definesymbol{\bedges}                           {boundary edges of \(\mesh\)}                            {boundaryedges}}
\newcommand{\defineFaces}                     {\definesymbol{\faces}                            {faces of \(\mesh\)}                                     {faces}}
\newcommand{\defineDualCell}                  {\definesymbol{\dualcell_i}                       {dual cell associated with vertex \(i\)}                 {dualcell}}
\newcommand{\defineCurve}                     {\definesymbol{\gamma(s)}                         {curve along boundary of domain}                         {curve}}
\newcommand{\defineNewCurve}                  {\definesymbol{\tgamma(\ts)}                      {new/target boundary curve}                              {newcurve}}
\newcommand{\defineGeodesicCurvature}         {\definesymbol{\kappa}                            {curvature of \(\gamma\)}                                {geodesiccurvature}}
\newcommand{\defineNewGeodesicCurvature}      {\definesymbol{\tkappa}                           {curvature of \(\tgamma\)}                               {newgeodesiccurvature}}
\newcommand{\defineArcLength}                 {\definesymbol{\s}                                {arc length along domain boundary}                       {arclength}}
\newcommand{\defineNewArcLength}              {\definesymbol{\ts}                               {arc length of new boundary}                             {newarclength}}
\newcommand{\defineLengthDensity}             {\definesymbol{\ds}                               {length density of domain boundary}                      {lengthdensity}}
\newcommand{\defineNewLengthDensity}          {\definesymbol{\tds}                              {length density of new boundary}                         {newlengthdensity}}
\newcommand{\defineAreaDensity}               {\definesymbol{\dA}                               {area density of domain}                                 {areadensity}}
\newcommand{\defineHilbertTransform}          {\definesymbol{\Htransform}                       {Hilbert transform}                                      {hilberttransform}}
\newcommand{\defineLogFactor}                 {\definesymbol{\u}                                {log conformal factor}                                   {logfactor}}
\newcommand{\defineRealPartF}                 {\definesymbol{\a}                                {real part of complex function}                          {realpartf}}
\newcommand{\defineImaginaryPartF}            {\definesymbol{\b}                                {imaginary part of complex function}                     {imaginarypartf}}
\newcommand{\defineAdjustFactor}              {\definesymbol{\r}                                {length rescaling for curve closure}                     {adjustfactor}}
\newcommand{\defineDirichletData}             {\definesymbol{\g: \bM \to \RR}                   {Dirichlet boundary data}                                {dirichletdata}}
\newcommand{\defineNeumannData}               {\definesymbol{\h: \bM \to \RR}                   {Neumann boundary data}                                  {neumanndata}}
\newcommand{\definePoissonRHS}                {\definesymbol{\phi: \M \to \RR}                  {source term in Poisson equation}                        {poissonrhs}}
\newcommand{\defineDualBoundaryEdge}          {\definesymbol{\dbedge_i}                         {dual boundary edge at vertex \(i\)}                     {dualboundaryedge}}
\newcommand{\defineConformalEnergy}           {\definesymbol{\conformalenergy}                  {least-squares conformal energy}                         {conformalenergy}}
\newcommand{\defineCotanLaplace}              {\definesymbol{\cotLaplace}                       {cotan-Laplace matrix}                                   {cotanlaplace}}
\newcommand{\defineCholeskyFactor}            {\definesymbol{\Lsf}                              {Cholesky factor of cotan matrix \(\cotLaplace\)}        {choleskyfactor}}
\newcommand{\defineDiscreteGaussCurvature}    {\definesymbol{\Omega: \vertices \to \RR}         {discrete Gauss curvature density}                       {discretegausscurvature}}
\newcommand{\defineDiscreteGeodesicCurvature} {\definesymbol{\ksf}                              {discrete geodesic curvature density}                       {discretegeodesiccurvature}}
\newcommand{\defineLaplaceBeltrami}           {\definesymbol{\Delta}                            {Laplace-Beltrami operator on domain}                    {laplacebeltrami}}
\newcommand{\defineDirichletToNeumann}        {\definesymbol{\Lambda}                           {Dirichlet-to-Neumann map}                               {dirichlettoneumann}}
\newcommand{\defineEdgeLength}                {\definesymbol{\edgelength{\ij}}                  {length of edge \(\ij\)}                                 {edgelength}}
\newcommand{\defineDualLength}                {\definesymbol{\duallength{i}}                    {length of dual boundary edge \(\dbedge_i\)}             {duallength}}
\newcommand{\defineTargetLength}              {\definesymbol{\targetlength{\ij}}                {desired length of edge \(\ij\)}                         {targetlength}}
\newcommand{\defineFinalLength}               {\definesymbol{\finallength{\ij}}                 {new length of edge \(\ij\)}                             {finallength}}
\newcommand{\defineTargetConeAngles}          {\definesymbol{\Theta}                            {target cone angles}                                     {targetconeangles}}
\newcommand{\defineDiscreteScaleFactor}       {\definesymbol{\usf_i}                            {discrete scale factor at boundary vertex \(i\)}         {discretescalefactor}}
\newcommand{\defineDiscreteNeumannData}       {\definesymbol{\hsf_i}                            {discrete Neumann data at boundary vertex \(i\)}         {discreteneumanndata}}
\newcommand{\defineDiscreteDirichletData}     {\definesymbol{\gsf_i}                            {discrete Dirichlet data at boundary vertex \(i\)}       {discretedirichletdata}}
\newcommand{\defineDiscreteTangent}           {\definesymbol{\T_{\ij}}                          {discrete tangent along boundary edge \(\ij\)}           {discretetangent}}
\newcommand{\defineDiscreteFlattening}        {\definesymbol{\fsf: \vertices \to \CC}           {discrete flattening}                                    {discretemap}}
\newcommand{\defineNewExteriorAngles}         {\definesymbol{\tksf: \bvertices \to \RR}         {target exterior angles}                                 {newexteriorangles}}
\newcommand{\defineTargetTangent}             {\definesymbol{\tTsf: \bedges \to \CC}            {target unit tangents}                                   {targettangent}}

\begin{document}
% Title portion
\title{Boundary First Flattening} 
\author{Rohan Sawhney}
\author{Keenan Crane}
\orcid{0000-0003-2772-7034}
\affiliation{%
  \institution{Carnegie Mellon University}
  \streetaddress{5000 Forbes Ave}
  \city{Pittsburgh}
  \state{PA}
  \postcode{15213}
  \country{USA}}

\renewcommand\shortauthors{Sawhney and Crane}

\begin{abstract}
   A \emph{conformal flattening} maps a curved surface to the plane without distorting angles---such maps have become a fundamental building block for problems in geometry processing, numerical simulation, and computational design.  Yet existing methods provide little direct control over the shape of the flattened domain, or else demand expensive nonlinear optimization.  \emph{Boundary first flattening (BFF)} is a linear method for conformal parameterization which is faster than traditional linear methods, yet provides control and quality comparable to sophisticated nonlinear schemes.  The key insight is that the boundary data for many conformal mapping problems can be efficiently constructed via the \emph{Cherrier formula} together with a pair of \emph{Poincar\'{e}-Steklov operators}; once the boundary is known, the map can be easily extended over the rest of the domain.  Since computation demands only a single factorization of the real Laplace matrix, the amortized cost is about 50x less than any previously published technique for boundary-controlled conformal flattening.  As a result, BFF opens the door to real-time editing or fast optimization of high-resolution maps, with direct control over boundary length or angle.  We show how this method can be used to construct maps with sharp corners, cone singularities, minimal area distortion, and uniformization over the unit disk; we also demonstrate for the first time how a surface can be conformally flattened directly onto any given target shape.
\end{abstract}




% We no longer use \terms command
% \terms{Design, Algorithms, Performance}


\thanks{} % TODO

\maketitle

\begin{figure}[h!]
   \centering
   \includegraphics{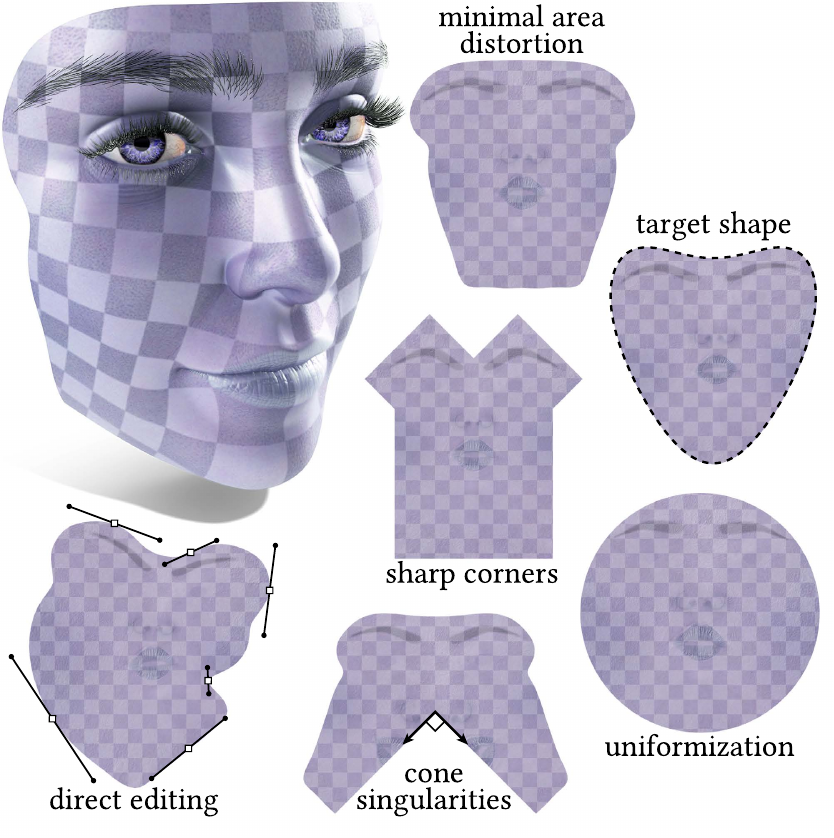}
   \caption{\emph{Boundary first flattening (BFF)} provides sophisticated control over the shape of an angle-preserving or \emph{conformal} map to the plane.  Even on very large meshes, this shape can be edited in real time.~\label{fig:teaser}}
\end{figure}

\section{Introduction}
\label{sec:Introduction}

In recent years conformal flattening has evolved far beyond its humble origins as a tool for texture mapping~\cite{Levy:2002:LSC,Desbrun:2002:IPS}, providing new perspectives on a broad range of applications including surface remeshing~\cite{Alliez:2003:ISR,Zhong:2014:ASM}, comparative data analysis~\cite{Hurdal:2009:DCM,Lipman:2011:CWD}, computational biology~\cite{Koehl:2015:LFG}, physical simulation~\cite{Segall:2016:HSF}, sensor networks~\cite{Rik:2009:GRG}, and computational design~\cite{Konakovic:2016:BDC}.  Why such great interest in maps that preserve \emph{angles}?  One answer is computational: conformal mapping typically amounts to solving easy linear equations, providing fast, scalable algorithms---or cheap initialization for more difficult nonlinear tasks~\cite{Liu:2008:LAM,Chao:2010:SGM}.  Another is that angle preservation is directly linked to real mechanical or constitutive properties of physical systems~\cite{Kim:2012:DRB}.  Continued advancement of basic tools for conformal mapping therefore has high utility across a variety of domains.  At present, however, linear algorithms for conformal flattening do not exploit the full space of possibilities.

In the smooth setting, conformal maps are quite flexible: a map from a disk-like surface to the flat plane can achieve any target shape as long as it is free to ``slide'' along the boundary.  In stark contrast, existing linear conformal flattening algorithms provide \emph{no} explicit control over the target shape: the user obtains a single, automatic solution and must ``take it or leave it.''  Nonlinear methods provide additional control over shape, but at significantly greater cost.  \emph{Boundary first flattening (BFF)} is the first conformal flattening method to provide full control over the target shape via a single sparse matrix factorization, including:
\begin{itemize}
   \item automatic flattening with optimal area distortion,
   \item direct manipulation of boundary length or angle,
   \item exact preservation of sharp corners,
   \item seamless cone parameterization,
   \item uniformization over the unit disk, and
   \item mapping to a given target shape.
\end{itemize}
(See examples in \figref{teaser}.)  The target shape can be efficiently updated via backsubstitution, providing a new paradigm for conformal parameterization: rather than waiting for a single, predetermined flattening, one can interactively edit or optimize the map---even on meshes with hundreds of thousands of elements.  More broadly, BFF is a drop-in replacement for widely-used schemes like least-squares conformal maps (LSCM) while providing sophisticated control over features like shape and area distortion.

\subsection{Algorithm Outline}
\label{sec:AlgorithmOutline}

Given a target length (or curvature) function along the boundary, the BFF algorithm involves three basic steps:
\begin{enumerate}[I.]
   \item Solve for a compatible curvature (or length) function.
   \item Integrate this data to get a boundary curve.
   \item Extend this curve over the interior of the domain.
\end{enumerate}
Computation amounts to solving three sparse linear problems (in steps I and III) connected by a simple nonlinear change of variables (step II); all linear problems use the same fixed Laplace matrix.  From here, the essential question is how to devise boundary data (curvatures or lengths) suitable for a variety of mapping problems, as explored in \secref{Applications}.  An important limitation of BFF is that it applies only to domains with disk topology, though of course a surface of any topology can be cut into one or more disks; see~\secref{GuaranteesAndLimitations} for further discussion.  Detailed pseudocode is provided in \appref{Pseudocode}.

\section{Related Work}
\label{sec:RelatedWork}

The literature on surface flattening and computational conformal geometry is vast~\cite{Sheffer:2006:MPM,Gu:2008:CCG}---here we focus on conformal flattening methods that provide control over boundary shape, or that can be modified to provide such control. Early methods for conformal flattening compute a piecewise linear least-squares solution of the \emph{Cauchy-Riemann equations} (\textbf{LSCM}~\cite{Levy:2002:LSC}), or equivalently, minimize the difference between \emph{Dirichlet energy} and the flattened area (\textbf{DCP}~\cite{Desbrun:2002:IPS});  quality was later improved via a sparse eigenvalue problem (\textbf{SCP}~\cite{Mullen:2008:SCP}).  A different line of methods directly optimizes angles in the flattened mesh to approximate angles in the input~\cite{Sheffer:2001:PFS,Sheffer:2004:ABF}, a process that can be linearized without compromising quality (\textbf{LinABF}~\cite{Zayer:2007:LAB}).

\setlength{\intextsep}{0pt}%
\setlength{\columnsep}{8pt}%
\begin{wrapfigure}{r}{104pt}
   \includegraphics{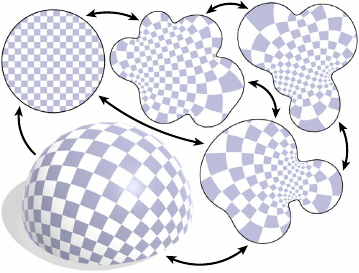}
\end{wrapfigure} 

\begin{figure}
   \centering
   \includegraphics{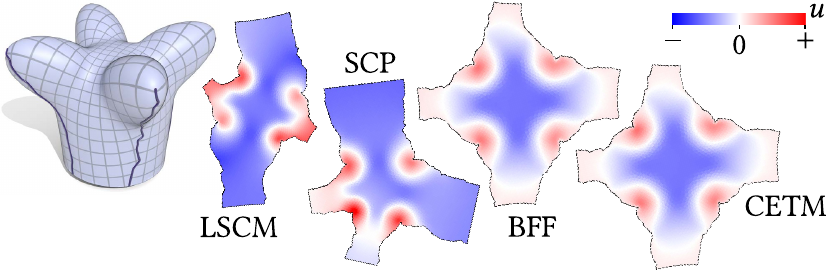}
   \caption{A surface is cut and conformally flattened (\(u\) is the scale factor).  In the absence of proper boundary conditions, traditional methods (LSCM and SCP) produce different solutions since the underlying smooth energy does not have a unique minimizer.  By adding explicit boundary conditions our method (BFF) provides a canonical map with minimal area distortion, nearly identical to more expensive nonlinear methods (CETM).\label{fig:AreaDistortion}}
\end{figure}

\paragraph{Free Boundary Conditions.} The methods mentioned so far forgo the question of boundary control, instead opting for so-called ``free'' boundary conditions where a discrete energy is minimized without explicit constraints on boundary degrees of freedom.  Although this approach provides a unique solution in the discrete setting, it has no meaningful interpretation in the smooth setting: in the absence of explicit boundary conditions there is an enormous space of perfect conformal flattenings, obtained by flattening and then applying an in-plane conformal map (see inset).  The only possibility is that \emph{the unique solution chosen by free boundary methods \textbf{must} depend on discretization}, meaning that results will change depending on the particular choice of mesh or numerical treatment (see \figref{AreaDistortion}, and \citet[Figure 1]{Mullen:2008:SCP}).  This phenomenon further motivates the need for more careful treatment of boundary conditions.

\paragraph{Prescribed Boundary Length/Angle.} Later methods explicitly incorporate boundary constraints.  For instance, the method of \emph{circle patterns} (\textbf{CP}~\cite{Kharevych:2006:DCM}) provides control over the direction of the boundary (but not its length); methods based on \emph{discrete Ricci flow}~\cite{Luo:2004:CYF,Jin:2008:DSR} can provide control over both length and direction (\textbf{CETM}~\cite{Springborn:2008:CET}).  All of these methods demand nonlinear optimization involving, \eg, repeated matrix factorization for each Newton iteration.  In \secref{ModifiedLinearMethods} we explore how linearization of both angle-based (LinABF) and length-based strategies (\textbf{CPMS}~\cite{BenChen:2008:CFC}) can be adapted to provide complete boundary control; even then, these methods remain at least 30x slower than BFF and can exhibit significant artifacts.  A very different approach is to discretize a \emph{time-independent Dirac equation} that governs conformal surface deformations in 3D~\citep[Section 6.1]{Crane:2013:CGP}; this approach provides control over boundary direction (but not length) and must solve an eigenvalue problem that cannot be prefactored for varying boundary data.

\paragraph{Uniformization.} A natural idea for achieving a given target shape is to compose maps to a canonical domain like the circular disk (\secref{Uniformization}).  In the discrete setting, however, piecewise linear conformal maps do not compose; more importantly, methods for computing such maps~\cite{Zeng:2008:SAC,Bobenko:2010:DCM} are already more expensive than just directly editing the boundary via BFF.

\paragraph{2D Shape Editing.} Methods for planar shape deformation are generally not suitable for conformal flattening since they depend on boundary element methods or closed-form expressions that are available only on planar domains~\cite{Weber:2010:CCM,Lipman:2012:SFF,Chen:2013:PSI}.  One idea is to apply 2D deformation to an initial flattening, but the resulting nonlinear map may be incompatible with the standard geometry processing pipeline, especially when one requires precise control over individual lengths and angles (\secref{Applications}).  Moreover, a composition of methods offers no clear advantage in terms of speed or simplicity over the unified scheme we propose here.

\mbox{}

Outside of strictly conformal methods a variety of algorithms provide boundary control, albeit with very different performance characteristics---see for example Weber \& Zorin~\shortcite{Weber:2014:LIP} and references therein.  At a high level, all previous boundary-controlled \emph{conformal} methods (namely, CP and CETM plus our modifications of CPMS and LinABF) indirectly encode a flattening via metric data (lengths or angles).  In order for this data to describe a valid planar triangulation it must satisfy nonlinear integrability conditions over the entire domain.  In contrast, BFF requires only that data describes a closed boundary \emph{curve}---a condition that is far easier to satisfy, once the associated integrability conditions are clearly understood.

% On the whole, BFF provides an attractive alternative to more sophisticated nonlinear methods like CETM, without compromising quality or controllability.

\section{Background}
\label{sec:Background}

We first provide some key definitions from (discrete) differential geometry; a more pedagogical introduction can be found in Crane \etal~\shortcite{Crane:2013:DGP} (\esp\ Chapter 7).  Throughout the document, clicking on most symbols will provide a hyperlink back to their definition.

\subsection{Notation}
\label{sec:Notation}

Single brackets denote the norm \(|\cdot|\) and real inner product \defineInnerVec \(\innervec{\cdot}{\cdot}\) of finite-dimensional vectors.  For any complex number \(z \in \CC\), \(\re(z)\) and \(\im(z)\) denote the real and imaginary parts, and the imaginary unit \defineImaginaryUnit \(\imath\) denotes a quarter-turn in the counter-clockwise direction (hence \(\imath^2 = -1\)).  The \defineArgument \emph{argument} \(\arg(z)\) of a complex number is the smallest (in magnitude) signed angle from the real axis to \(z\); the angle from \(z_1\) to \(z_2\) can therefore be expressed as \(\arg(z_1^{-1}z_2)\).  Italic glyphs (\(A,b,\ldots\)) typically indicate a continuous quantity whereas sans-serif characters (\(\Asf,\bsf\)) denote discrete quantities and/or matrices.

\subsection{Smooth Setting}
\label{sec:SmoothSetting}

\begin{figure}
   \centering
   \includegraphics{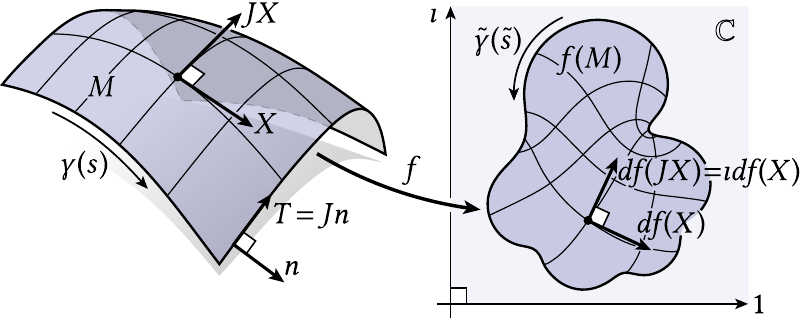}
   \caption{Basic quantities used in our algorithm, which computes a conformal map \(\f\) from a surface \(\M\) to the plane \(\CC\).~\label{fig:SmoothQuantities}}
\end{figure}

Our main object of study is a map \defineFlattening \(\f: \M \to \CC\) from a disk-like surface \defineSurface \(\M\) (with Riemannian metric) to the flat complex plane \(\CC\) (\figref{SmoothQuantities}). At each point the map \defineComplexStructure \(\J\) rotates any tangent vector \(X\) by a quarter-turn in the counter-clockwise direction so that (like the imaginary unit) \(\J^2 X = -X\).  The boundary \defineBoundary \(\bM\) is a single closed loop parameterized by a curve \defineCurve \(\gamma\) with arc-length parameter \defineArcLength \(\s\).  We likewise use \defineNewCurve \defineNewArcLength \(\tgamma(\ts)\) to parameterize the image of the boundary \(\f(\bM)\); it is the shape of this curve that we seek to control.  By convention, if \defineTangent \(\T\) is the unit tangent in the direction \(\tfrac{d}{d\s} \gamma\), then \defineNormal \(\n := -\J \T\) is the outward unit normal.  We use \defineGaussCurvature \(\K\) to denote the Gaussian curvature of \(\M\), and \defineGeodesicCurvature\defineNewGeodesicCurvature \(\kappa, \tkappa\) for the (geodesic) curvatures of \(\gamma\) and \(\tgamma\), respectively.

\subsubsection{Conformal Maps}
\label{sec:ConformalMaps}

Intuitively, a map \(\f: \M \to \CC\) is conformal if at each point it preserves the angle between any two vectors, permitting only a uniform change in length.  More precisely, let \defineDifferential \(\df\) denote the \emph{differential} of \(\f\), which determines how a given tangent vector \(X\) on the surface gets mapped to a tangent vector \(\df(X)\) in the complex plane (in coordinates, \(\df\) is represented by the familiar \emph{Jacobian matrix}).  A map \(\f\) is \emph{holomorphic} if
\begin{equation}
   \label{eq:CauchyRiemann}
   \df(\J X) = \imath \df(X)
\end{equation}
for all tangent vectors \(X\), \ie, if a quarter-turn \(\J\) on the surface yields the same result as a quarter-turn \(\imath\) in the plane; this relationship is known as the \emph{Cauchy-Riemann equation}.  If in addition \(\df\) is nondegenerate (\ie, it maps nonzero vectors to nonzero vectors) then \(\f\) is \emph{conformal}.  The \emph{conformal (scale) factor} \(e^{\u} := |\df(X)|/|X|\) quantifies the change in length at each point (which is independent of direction \(X\)); the function \defineLogFactor \(\u: \M \to \RR\) is called the \emph{log conformal factor}.

Conformal maps can also be expressed as pairs of \emph{conjugate harmonic functions}.  A real function \(a: \M \to \RR\) is \emph{harmonic} if it solves the \emph{Laplace equation} \(\Delta a = 0\) (\secref{PoissonProblems}), where \(\Delta\) is the \defineLaplaceBeltrami \emph{Laplace-Beltrami operator} (or just \emph{Laplacian}) associated with the domain \(\M\).  Suppose we express a holomorphic map as \defineRealPartF \defineImaginaryPartF \(\f = \a + \b \imath\) for a pair of coordinate functions \(\a,\b: \M \to \RR\).  Then (by Cauchy-Riemann)
\begin{equation}
   \label{eq:ConjugateHarmonic}
   J \nabla a = \nabla b,
\end{equation}
\ie, the gradients \(\nabla\) of the two coordinates are orthogonal and have equal magnitude.  Since a quarter-rotation of a gradient field is divergence-free, we have
\[
   \begin{array}{rcrcrcl}
      \Delta a &=& \nabla \cdot \nabla a &=& -\nabla \cdot (J\nabla b) &=& 0,
   \end{array}
\]
and similarly, \(\Delta b = 0\).  In other words, the two real components of a holomorphic function are both \emph{harmonic}---we say that \(a\) and \(b\) form a \emph{conjugate harmonic} pair.

\begin{figure}[b]
   \centering
   \includegraphics{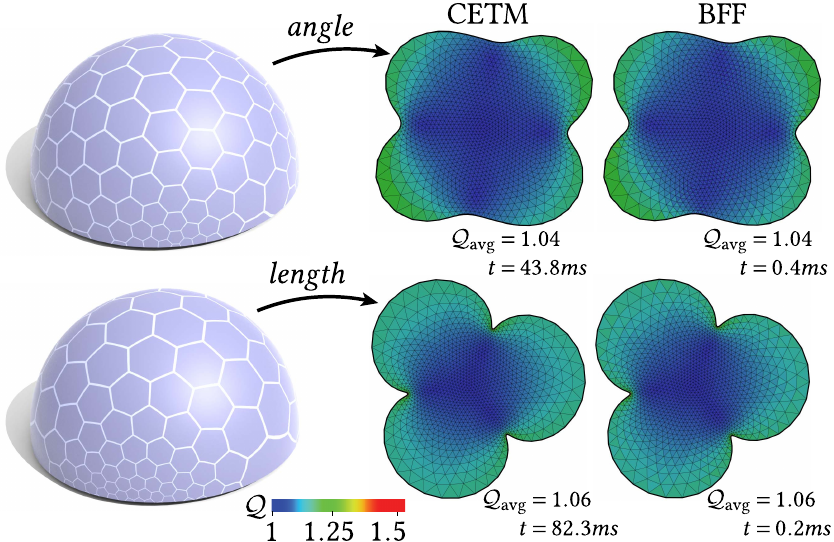}
   \caption{A conformal flattening is uniquely determined (up to global similarity) by either length \figloc{(bottom)} or curvature \figloc{(top)} data along the boundary.  Here we prescribe simple oscillatory functions, achieving results virtually indistinguishable from a reference solution (CETM) despite the fact that BFF is over 100x faster. (Hexagonal pattern emphasizes preservation of angles.)\label{fig:OscillatoryAngleLength}}
\end{figure}

To what degree can one control the target shape of a conformal flattening?  Equation~\ref{eq:ConjugateHarmonic} suggests there must be some limitation, since two arbitrary functions on the boundary may not extend to a conjugate harmonic pair on the interior.  And yet the \emph{Riemann mapping theorem} states that any disk-like region \(U \subset \CC\) can be conformally mapped to the unit circular disk (\secref{Uniformization}); hence any shape can be obtained via composition with an (inverse) Riemann map.  These two facts do not contradict each-other: although one can surjectively map onto any given region (\(\f(\M) = U\)), one cannot arbitrarily ``pin'' each boundary point to a specific location (\(\f|_{\bM} = \tgamma\)) and hope to remain conformal.
%(see for example \figref{HarmonicVsConformal})
We therefore express our basic algorithm in terms of geometric quantities that can always be realized: one can specify \emph{either} length \emph{or} curvature density along the boundary, \emph{but not both simultaneously} (see \figref{OscillatoryAngleLength}, \secref{CherrierFormula} and \citep[Section 4]{Springborn:2008:CET}).  This data can then be further manipulated to provide a variety of intuitive control schemes (\secref{Applications}).

\subsubsection{Curvature Density}
\label{sec:CurvatureDensity}

In developing boundary conditions it helps to distinguish between the usual pointwise curvatures \(\kappa\), \(\K\), and the corresponding curvature \emph{densities}, because of the way these quantities transform under a conformal map.  Loosely speaking, a \emph{density} assigns a positive volume to each little piece of a curve or surface, \eg, the length density \defineLengthDensity \(\ds\) on a curve or the area density \defineAreaDensity \(\dA\) on a surface.  Under a conformal map, new and old densities scale according to the conformal factor: \defineNewLengthDensity \(\tds = e^{\u} \ds\) and \(\dtA = e^{2\u} \dA\).  A \emph{curvature density} multiples a density by a curvature function---in particular, \(\kappa \ds\) is the (geodesic) curvature density of a curve, and \(\K \dA\) is the Gauss curvature density of a surface.  The change of curvature density under a conformal map is studied in \secref{CherrierFormula}.

\subsection{Discrete Setting}
\label{sec:DiscreteSetting}

\setlength{\intextsep}{0pt}%
\setlength{\columnsep}{8pt}%
\begin{wrapfigure}{l}{119pt}
   \centering
   \includegraphics{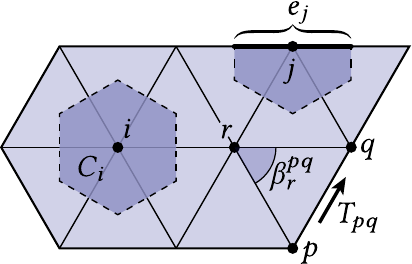}
\end{wrapfigure}
We discretize our surface \(M\) as a manifold triangle mesh \defineMesh\defineVertices\defineEdges\defineFaces \(\mesh = (\vertices,\edges,\faces)\) with disk topology, using \defineBoundaryVertices \(\bvertices \subseteq \vertices\) to denote the set of vertices on the boundary \defineBoundaryEdges \(\bedges\), and \defineInteriorVertices \(\ivertices := \vertices \setminus \bvertices\) for interior vertices.  Tuples of vertex indices are used to specify simplices, \eg, \(\ij \in \edges\) is an edge from vertex \(i\) to vertex \(j\).  An expression of the form \(a_i = \sum_{\ij \in \edges} b_{\ij}\) means that a quantity \(b\) is summed over all edges containing vertex \(i\) to get the value of \(a\) at \(i\).  Likewise, \(a_i = \sum_{\ijk \in \faces} b_{\ijk}\) denotes a sum over faces containing vertex \(i\).  A quantity at corner \(i\) of triangle \(\ijk\) is denoted by a subscript \(i\) and superscript \(jk\); for instance, we use \defineTipAngle \smash{\(\tipangle_i^{jk} \in \RR\)} to denote the interior angle at the corner of a triangle.  Throughout we use \defineDualCell \(\dualcell_i\) to denote the dual cell associated with vertex \(i \in \vertices\), and \defineDualBoundaryEdge \(\dbedge_j\) to denote the barycentric dual edge associated with a boundary vertex \(j \in \bvertices\); \defineDiscreteTangent \(\Tsf_{pq}\) denotes the unit tangent along boundary edge \(pq\) (see inset).  We use \defineEdgeLength \(\edgelength{\ij}\) to denote the length of edge \(\ij\) in the input mesh, and \defineDualLength \(\duallength{j} := \tfrac{1}{2}(\edgelength{\ij} + \edgelength{\jk})\) to denote the length of the dual boundary edge \(\dbedge_j\), where \(i\), \(j\), and \(k\) are consecutive vertices along the boundary.

\subsubsection{Discrete Curvature}
\label{sec:DiscreteCurvature}

\setlength{\intextsep}{0pt}%
\setlength{\columnsep}{8pt}%
\begin{wrapfigure}{r}{138pt}
   \centering
   \includegraphics{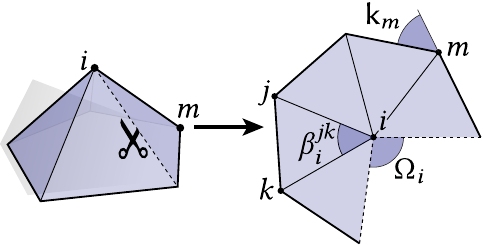}
\end{wrapfigure}
\paragraph{Gaussian curvature.} For each interior vertex \(i\) of a triangulation the angle defect \defineDiscreteGaussCurvature \(\Omega_i := 2\pi - \sum_{ijk \in \faces} \tipangle_i^{jk}\) quantifies the ``flatness'' of a vertex as a deviation from the planar angle sum of \(2\pi\).  This quantity encodes not the \emph{pointwise} Gaussian curvature, but rather the curvature integrated over a small region around the vertex: \(\Omega_i = \int_{\dualcell_i} \K\ \dA\) (since for a discrete surface \(K\) is a distribution concentrated at vertices).  We therefore refer to \(\Omega\) as a \emph{discrete curvature density}.

\paragraph{Geodesic curvature.} At boundary vertices, the Gaussian curvature (density) \(\Omega_i\) is zero, since a small neighborhood around any boundary vertex \(i \in \bvertices\) can be flattened into the plane without stretching.  Here we consider a different quantity \defineDiscreteGeodesicCurvature \(\ksf_i := \pi - \sum_{ijk \in \faces} \smash{\tipangle_i^{jk}}\), which encodes the integral of the \emph{geodesic} curvature \(\kappa\) (again a distribution at vertices).  In the plane, \(\ksf_i\) are also the \emph{exterior angles}, \ie, the change in tangent direction from one edge to the next.

\setlength{\intextsep}{0pt}%
\setlength{\columnsep}{8pt}%
\begin{wrapfigure}{r}{88pt}
   \centering
   \includegraphics{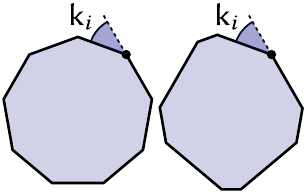}
\end{wrapfigure}
The discrete setting helps illustrate the challenge of prescribing pointwise curvature rather than curvature density.  Suppose we want to achieve a boundary curvature \(\kappa\).  If we provide only exterior angles \(\ksf\), then there are in general many polygons that agree with this data (see inset).  To uniquely prescribe a target shape, we must be able to control the change in angle \emph{per unit length}.  Likewise, to prescribe Gaussian curvature, we must be able to control the angle defect \emph{per unit area}.  Algorithmically this goal is difficult to achieve since the final lengths and areas are are determined by a scale factor \(\u\) that is not known \emph{a priori}---any algorithm that prescribes angle defects (such as CPMS or CETM) actually controls the \emph{(discrete) curvature density} rather than the curvature itself.  One approach to this problem is described in \secref{ArbitraryCurves}.

\section{Toolbox}
\label{sec:Toolbox}

We first describe a collection of basic building blocks, which are assembled into the final algorithm in \secref{Algorithm}.

\subsection{Poisson Problems}
\label{sec:PoissonProblems}

The development of BFF hinges on careful treatment of boundary conditions for the \emph{Dirichlet-Poisson problem}
\begin{equation}
   \label{eq:DirichletPoisson}
      \Delta \a = \phi \quad \on \M; \qquad \a = \g \quad \on \bM
\end{equation}
and the \emph{Neumann-Poisson problem}
\begin{equation}
   \label{eq:NeumannPoisson}
      \Delta \a = \phi \quad \on \M; \qquad \tfrac{\partial \a}{\partial \n} = \h \quad \on \bM,
\end{equation}
where \(\a\) and \definePoissonRHS \(\phi\) are real-valued functions on \(\M\), and \defineDirichletData \defineNeumannData \(\g,\h: \partial \M \to \mathbb{R}\) determine the values or normal derivatives along the boundary, \resp\phantom{.}  The solution to \eqref{NeumannPoisson} is unique only up to a constant, which in BFF just determines the global scale and translation.  On a triangle mesh, integrating \(\Delta \a = \phi\) over dual cells yields a matrix equation
\[
   \cotLaplace \asf = \Psf \phi,
\]
where \defineCotanLaplace \(\cotLaplace \in \mathbb{R}^{|\vertices| \times |\vertices|}\) is the so-called \emph{cotan-Laplace matrix} since its nonzero entries can be expressed as
\[
   \cotLaplace_{\ij} = -\tfrac{1}{2}( \cot\tipangle_p^{\ij} + \cot\tipangle_q^{\ij} )
\]
for each edge \(\ij \in \edges\) with opposite vertices \(p,q\), and
\(
   \cotLaplace_{\ii} = -\sum_{\ij \in \edges} \cotLaplace_{\ij}
\)
for each vertex \(i \in \vertices\) (see \citet[Section 6.3]{Crane:2013:DGP} for a derivation); omitting the unknown cotans at boundary edges corresponds to zero-Neumann boundary conditions (\citet[Eqn. 3.3]{MacNeal:1949:SPD}).  The matrix \(\Psf\) is the \emph{mass matrix}, but is not needed for BFF since all values appearing on the right-hand side will be integrated quantities (\eg, the discrete Gaussian curvature \(\Omega\)).

If we partition into interior vertices \(\ivertices\) and boundary vertices \(\bvertices\), a Neumann-Poisson problem can be expressed in block form as
\begin{equation}
   \label{eq:DiscretePoissonNeumann}
   \left[ \begin{array}{ll} \cotLaplace_{\II} & \cotLaplace_{\IB} \\ \cotLaplace_{\IB}^\transpose & \cotLaplace_{\BB} \end{array} \right] \left[ \begin{array}{c} \asf_I \\ \asf_B \end{array} \right] = \left[ \begin{array}{l} \phi_{\ivertices} \\ \phi_B-\hsf \end{array} \right],
\end{equation}
where \defineDiscreteNeumannData \(\hsf \in \RR^{|\bvertices|}\) is the discrete Neumann boundary data, corresponding to the \emph{integral} of \(\partial \a/\partial \n\) over each dual boundary edge \(\dbedge_i\).  A Dirichlet problem with boundary values \defineDiscreteDirichletData \(\asf_B = \gsf \in \RR^{|\bvertices|}\) is then obtained by solving the first row of \eqref{DiscretePoissonNeumann} for the interior values \(\asf_I\):
\begin{equation}
   \label{eq:DiscretePoissonDirichlet}
   \cotLaplace_{\II} \asf_I = \phi_I - \cotLaplace_{\IB} \gsf.
\end{equation}

\paragraph{Efficient Solution} Solving a sequence of Dirichlet- and Neumann-Poisson problems requires only a single sparse Cholesky factorization \defineCholeskyFactor \(\cotLaplace = \Lsf\Lsf^\transpose\) of the matrix \(\cotLaplace\) from the Neumann problem.  To see why, consider the blockwise expansion
\begin{equation}
   \label{eq:BlockFactorization}
\def\arraystretch{1.1}
\left[ \begin{array}{ll} \cotLaplace_{\II} & \cotLaplace_{\IB} \\ \cotLaplace_{\IB}^\transpose & \cotLaplace_{\BB} \end{array} \right] = \left[ \begin{array}{ll} \Lsf_{\II} & 0 \\ \Lsf_{\BI} & \Lsf_{\BB} \end{array} \right] \left[ \begin{array}{ll} \Lsf_{\II}^\transpose & \Lsf_{\BI}^\transpose \\ 0 & \Lsf_{\BB}^\transpose \end{array} \right],
\end{equation}
which means the Cholesky factorization for the Dirichlet problem is already given by the upper-left block: \(\cotLaplace_{\II} = \Lsf_{\II} \Lsf_{\II}^\transpose\).  All subsequent problems can then be solved (via backsubstitution) at a small fraction of the factorization cost---since BFF solves \emph{only} Poisson problems, the computational bottleneck is the single factorization of \(\cotLaplace\).  (In practice one needs to be careful about reordering; see \secref{QualityAndPerformance} for details.)  Note that no corresponding treatment is known for LSCM/SCP despite repeated \(\cotLaplace\) blocks---the closest proposal entails either dense factorization, or else iterative solvers with no amortized gains from prefactorization~\citep[Section 5.2]{Alexa:2011:DLG}.

\subsection{Cherrier Formula}
\label{sec:CherrierFormula}

The change in curvature under a conformal mapping has a close relationship with the scale factor \(\u\).  For domains without boundary this relationship is captured by the \emph{Yamabe problem}~\cite[Chapter 5]{Aubin:1998:SNP}, but our method depends critically on additional boundary conditions studied by Pascal Cherrier~\shortcite{Cherrier:1984:PNN}.  In particular, for a conformal map \(\f: \M \to \smash{\widetilde{M}}\) between any two surfaces,
\begin{equation}
   \label{eq:CherrierFormula}
   \begin{array}{rccl}
      \Delta \u &=& \K - e^{2\u} \widetilde{K} & \on \M \\
      \tfrac{\partial \u}{\partial \n} &=& \kappa - e^{\u} \tkappa & \on \bM
   \end{array}
\end{equation}
where \(\Delta\) is the Laplacian on \(\M\), and \(\widetilde{K},\tkappa\) are the new curvatures on \(\widetilde{M}\).  Although the equation \(\Delta u = \K - e^{2\u} \widetilde{K}\) is standard, its boundary conditions have been largely neglected by conformal flattening algorithms, which instead optimize discrete variables without reference to an underlying continuous equation~\cite{Kharevych:2006:DCM,Springborn:2008:CET}, or consider only the special case where \(\tkappa\) is a delta distribution~\cite{Bunin:2008:CTU,Myles:2013:CCG}.  We address the general case by multiplying \eqref{CherrierFormula} through by \(\dA\) and \(\ds\) (\resp) to obtain a linear relationship between \emph{densities}:
\begin{equation}
   \label{eq:GaussCurvatureDensityChange}
    \Delta \u\ \dA = \K \dA - \widetilde{K} \dtA,
\end{equation}
\begin{equation}
   \label{eq:GeodesicCurvatureDensityChange}
   \tfrac{\partial \u}{\partial \n} \ds = \kappa \ds - \tkappa \tds.
\end{equation}
Integrating \eqref{GaussCurvatureDensityChange} over dual cells hence yields a linear relationship \defineDiscreteScaleFactor
\begin{equation}
   \label{eq:DiscreteGaussCurvatureChange}
   \cotLaplace \usf = \Omega - \widetilde{\Omega},
\end{equation}
where \(\Omega, \widetilde{\Omega}: \vertices \to \RR\) are the source and target angle defects (see Sections~\ref{sec:DiscreteCurvature} and \ref{sec:PoissonProblems}); for flattening, \(\widetilde{\Omega} = 0\).  Likewise, integrating \eqref{GeodesicCurvatureDensityChange} over dual boundary edges \(\dbedge_i\) yields the linear equation
\begin{equation}
   \label{eq:DiscreteCurvatureChange}
   \hsf = \ksf - \tksf,
\end{equation}
where \(\hsf\) is the discrete Neumann data, and \defineNewExteriorAngles \(\ksf,\tksf\) are discrete boundary curvatures on the domain and target.  Note that (as discussed in Ben-Chen \etal~\citeyear[Section 2.3]{BenChen:2008:CFC}) one does \emph{not} obtain exact target curvatures if \(\usf\) values are used to rescale edge lengths (\ala\ CETM); in BFF we take an alternate route where angles are obtained directly from the discrete Cherrier formula, and exactly satisfy necessary conditions for closure of the target boundary loop (Proposition 1) and realization of target cone angles (Proposition 2).

\subsection{Poincar\'{e}-Steklov Operators}
\label{sec:PoincareSteklovOperators}

A given solution to an elliptic boundary value problem (like Poisson) can often be explained by several different types of boundary conditions---for instance, a harmonic function is uniquely determined by either its values (Dirichlet) or normal derivatives (Neumann) along the boundary.  In general, a \emph{Poincar\'{e}-Steklov operator} maps boundary data from one solution to alternative boundary data that yields an identical solution.  We require two such operators: the \emph{Dirichlet-to-Neumann map} for a Poisson equation, and the \emph{Hilbert transform} for the Cauchy-Riemann equation.

\subsubsection{Dirichlet to Neumann}
\label{sec:DirichletToNeumann}

\begin{figure}
   \centering
   \includegraphics{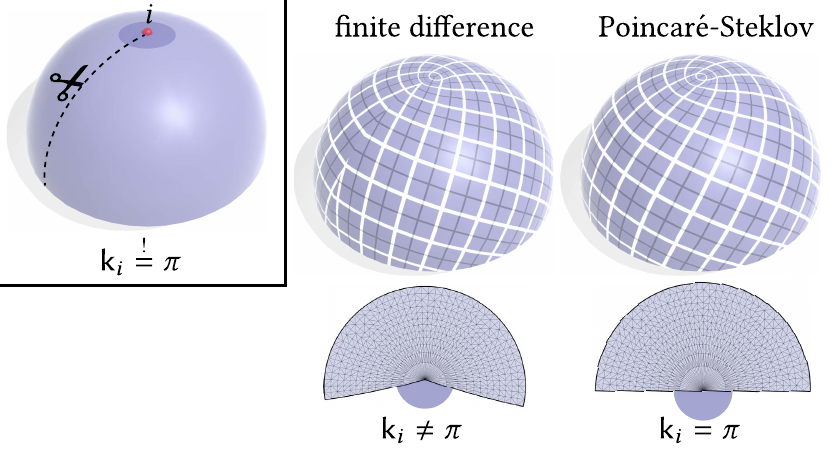}
   \caption{Accurate evaluation of boundary data is essential for achieving the correct boundary shape.  Here we attempt to prescribe an angle \(\pi\) at vertex \(i\) of a cut hemisphere using either a simple finite difference formula \figloc{(left)} or a Poincar\'{e}-Steklov operator \figloc{(right)} to obtain Neumann data; only the latter scheme yields the correct angle, producing a seamless map.\label{fig:SteklovVsGradDotN}}
\end{figure}

Given Dirichlet boundary values \(\gsf\) for a discrete Poisson equation, we seek Neumann values \(\hsf\) that yield the same solution.  One idea is to simply solve the Dirichlet problem, then evaluate the normal derivative directly---\eg, dot unit normals with the solution gradient, then integrate over dual boundary edges \(\dbedge_i\).  In practice, however, this approach yields poor behavior (see \figref{SteklovVsGradDotN}, \figloc{center}).  A more principled approach is to solve for Neumann data \(\hsf\) that exactly reproduces the discrete Dirichlet solution.  For a Laplace problem (\ie, \(\phi = 0\)) solving Equations \ref{eq:DiscretePoissonNeumann} and \ref{eq:DiscretePoissonDirichlet} for \(\hsf\) entails evaluating the so-called \emph{Schur complement} of \(\cotLaplace\):
\begin{equation}
   \label{eq:DirichletToNeumannLaplace}
   \hsf = (\cotLaplace_{\IB}^\transpose \cotLaplace_{\II}^{-1}\cotLaplace_{\IB} - \cotLaplace_{\BB}) \gsf.
\end{equation}
For a Poisson equation with nonzero source term \(\phi\), the \defineDirichletToNeumann Dirichlet-to-Neumann map becomes an \emph{affine} operator given by the expression
\begin{equation}
   \label{eq:DirichletToNeumannPoisson}
    \Lambda_\phi \gsf := \phi_B - \cotLaplace_\IB^\transpose \cotLaplace_\II^{-1} ( \phi_I - \cotLaplace_\IB \gsf ) - \cotLaplace_\BB \gsf.
\end{equation}
In practice this map can be evaluted via a single linear solve involving the (prefactored) matrix \(\cotLaplace_\II\) together with a sequence of basic matrix operations, as detailed in \algref{DirichletToNeumann}.  In effect, \(\Lambda_\phi\) solves a Dirichlet-Poisson equation, then takes the difference between the source term \(\phi\) and the Laplacian of the Poisson solution at each boundary node.  Beyond improved numerical behavior (\figref{SteklovVsGradDotN}, \figloc{right}), this discretization provides a useful invariant in the context of conformal flattening: discrete curvatures \(\ksf\) computed via \eqref{DiscreteCurvatureChange} always sum to \emph{exactly} \(2\pi\) (\appref{ExactAngleSums}), automatically satisfying a necessary condition for integrability (\secref{CurveIntegration}).

\subsubsection{Neumann to Dirichlet}
\label{sec:NeumannToDirichlet}

The reverse direction is more straightforward: one can simply solve the Neumann-Poisson equation (\eqref{DiscretePoissonNeumann}) for \(\asf\) and read off the boundary values \(\gsf := \asf_B\) (\algref{NeumannToDirichlet}).  Since the solution is determined only up to a constant (\secref{PoissonProblems}), we denote the Neumann-to-Dirichlet map by the pseudoinverse \(\Lambda^{\dagger}_\phi\).

\defineHilbertTransform 

\subsubsection{Hilbert Transform}
\label{sec:HilbertTransform}

On a disk-like domain, the \emph{Hilbert transform} \(\Htransform\) maps the tangential derivative of a harmonic function \(\a\) to the normal derivative of its harmonic conjugate \(\b\), providing boundary data for a holomorphic function \(\f = \a + \b\imath\) (\secref{ConformalMaps}).  Notions of conjugacy have been studied for a wide variety of discretizations~\cite{Polthier:2000:CHM,Mercat:2001:DRS,Weber:2010:CCM,Bobenko:2015:DCA}; in our context, we seek a pair of standard piecewise linear functions with degrees of freedom at vertices.  The basic idea is to fix \(\a\) and solve for the function \(\b\) that minimizes the \emph{least-squares conformal energy} \defineConformalEnergy \(\conformalenergy\), which measures the failure of \(\f\) to satisfy Cauchy-Riemann~\cite{Levy:2002:LSC}.  This energy can also be expressed as the Dirichlet energy of \(\f\) minus the area of its image~\cite{Hutchinson:1991:CCM}, which in matrix form becomes
\begin{equation}
   \label{eq:DiscreteConformalEnergy}
   \conformalenergy(\asf,\bsf) = \left[ \begin{array}{cc} \asf^\transpose & \bsf^\transpose \end{array} \right] \left[ \begin{array}{cc} \cotLaplace & \Usf \\ \Usf^\transpose & \cotLaplace \end{array} \right] \left[ \begin{array}{c} \asf \\ \bsf \end{array} \right],
\end{equation}
where \(\Usf\) encodes the signed area of the boundary polygon:
\begin{equation}
   \label{eq:SignedArea}
   \asf^\transpose \Usf \bsf := \tfrac{1}{2} \sum_{\ij \in \bedges} \asf_j \bsf_i - \asf_i \bsf_j
\end{equation}
(see \cite[Section 2.2]{Mullen:2008:SCP}).  Minimizing \(\conformalenergy\) with fixed \(\a\) therefore amounts to solving the Neumann-Laplace equation \( \cotLaplace \bsf = -\Usf^\transpose \asf \).  The Neumann boundary data \(\hsf = \Usf^\transpose \asf\) can be obtained by differentiating \eqref{SignedArea} with respect to \(\b\), yielding the simple expression
\begin{equation}
   \label{eq:HilbertBoundaryData}
   \hsf_j := \tfrac{1}{2} ( \asf_k - \asf_i ),
\end{equation}
for any three consecutive vertices \(i,j,k\) along the boundary.  Notably, this expression looks like a na\"{i}ve finite difference---the variational interpretation above verifies that it nonetheless yields a solution that is ``as conjugate as possible'' in the least-squares sense.

\subsection{Interpolation}
\label{sec:Interpolation}

Suppose we want to extend a given function \(\tgamma: \bM \to \CC\) over the interior of the domain, \ie, find a map \(\f: \M \to \CC\) such that \(\f|_{\bM} = \tgamma\).  A simple strategy is to independently interpolate each coordinate of \(\tgamma\) by a harmonic function, \ie, solve a pair of Laplace problems
\[
   \begin{array}{rcl}
      \Delta \a = 0 & \mathrm{s.t.} & \a|_{\bM} = \mathrm{Re}(\tgamma), \\
      \Delta \b = 0 & \mathrm{s.t.} & \b|_{\bM} = \mathrm{Im}(\tgamma).
   \end{array}
\]
If \(\tgamma\) is already compatible with some holomorphic map \(\f\), the Hilbert transform \(\Htransform\) from the previous section provides a different strategy:
\begin{enumerate}
   \item \emph{(Harmonic extension.)} Solve \(\Delta \a = 0\) s.t. \(\a|_{\bM} = \mathrm{Re}(\tgamma)\).
   \item \emph{(Harmonic conjugation.)} Solve \(\Delta \b = 0\) s.t. \(\tfrac{\partial \b}{\partial \n} = \Htransform \a\),
\end{enumerate}
where the discrete Neumann data for the latter problem is computed via \eqref{HilbertBoundaryData}.  These two strategies provide different algorithmic invariants: with the former, \(\f\) exactly interpolates \(\tgamma\) but is not guaranteed to be exactly holomorphic; with the latter \(\f\) is holomorphic but may not exactly interpolate both components of \(\tgamma\).  When \(\tgamma\) comes from a holomorphic map they will coincide; in practice, each is best-suited to different applications, as discussed in \secref{Applications}.

\begin{figure}[t]
   \centering
   \includegraphics{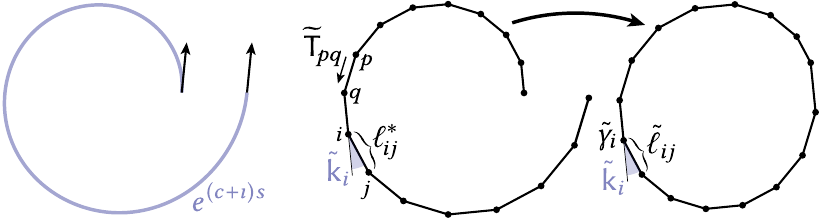}
   \caption{Simply ensuring that curvature integrates to \(2\pi\) is insufficient to guarantee that a boundary loop closes \figloc{(left)}.  We minimally adjust the desired edge lengths \(\targetlength{}\) to obtain a nearby closed loop with identical angles \(\tksf\) \figloc{(right)}.\label{fig:CurveClosure}}
\end{figure}

\subsection{Curve Integration}
\label{sec:CurveIntegration}

A key step in BFF is recovering a closed boundary curve \(\tgamma\) from given curvature and length data.  In the smooth setting this data can be integrated directly, but in the discrete case a small amount of discretization error prevents closure---we therefore seek a closed curve that approximates the given data.  Directly optimizing vertex positions leads to a difficult nonlinear problem; we instead solve an easy convex problem that closes the curve by minimally adjusting length.  In the smooth setting, we formulate this problem by considering the cumulative curvature
\[
   \varphi(t) := \int_0^t \kappa(\s)\ \ds.
\]
We then construct unit tangents \(\T(\s) := e^{\imath\varphi(\s)}\) along the boundary and solve the problem \defineAdjustFactor
\begin{equation}
   \label{eq:LengthMatching}
      \displaystyle\min_{\r: \bM \to \RR} \tfrac{1}{2} \int_0^{2\pi} \!\!\!(\r(\s) - 1)^2\ \ds \quad
      \mathrm{s.t.} \quad \int_0^{2\pi} \!\!\!\r(\s)\T(\s)\ \ds = 0
\end{equation}
If \(\kappa\) already describes a closed loop then we recover the solution \(\r \equiv 1\); otherwise, \(\r\) minimally adjusts the speed of the curve such that it closes.  In either case, we obtain the final curve by integrating the scaled tangents:
\(
   \tgamma(t) := \int_0^t \r(\s) \T(\s)\ \ds.
\)

\subsubsection{Discretization}
\label{sec:IntegrationDiscretization}

To discretize \eqref{LengthMatching}, let \(\tksf, \defineTargetLength \targetlength{}\) specify the desired exterior angles and edge lengths (\figref{CurveClosure}, \figloc{right}).  We seek a polygon with vertices \(\tgamma_i \in \CC\) that exactly achieves these angles and closely matches the lengths.  We first compute cumulative angles 
\[
   \varphi_p := \sum_{i=1}^{p-1} \tilde{\kappa}_i
\]
and target tangents \defineTargetTangent \(\tTsf_{\ij} := (\cos\varphi_i,\sin\varphi_i)\).  \eqref{LengthMatching} then becomes \defineFinalLength
\begin{equation}
   \label{eq:DiscreteBestFitLengths}
   \min_{\finallength{}: \bvertices \to \RR} \tfrac{1}{2}\!\!\sum_{\ij \in \bedges} \edgelength{\ij}^{-1} |\finallength{\ij} - \targetlength{\ij}|^2\ \quad\ \mathrm{s.t.}\ \quad\ \sum_{\ij \in \bedges} \finallength{\ij} \tTsf_{\ij} = 0.
\end{equation}
If \(\Nsf \in \RR^{|\bvertices| \times |\bvertices|}\) is a diagonal mass matrix with entries \(\Nsf_{\ii} = 1/\edgelength{\ij}\) and we pack the two coordinates of each unit tangent into a matrix \(\tTsf \in \RR^{2 \times |\bvertices|}\), then the optimal lengths are given by
\begin{equation}
   \label{eq:OptimalLengths}
   \finallength{} = \targetlength{} - \Nsf^{-1}\tTsf^\transpose(\tTsf\Nsf^{-1}\tTsf^\transpose)^{-1}\tTsf \targetlength{}.
\end{equation}
Note that \(\tTsf\Nsf^{-1}\tTsf^\transpose\) is just a \(2 \times 2\) matrix which costs virtually nothing to build and invert.  Final vertex positions are recovered via the cumulative sums
\(
   \smash{ \tgamma_p := \sum_{i=1}^{p-1} \finallength{\ij} \tTsf_{\ij}. }
\)
In principle the new lengths \(\finallength{\ij}\) could become negative, but in practice we do not observe this behavior: typical values for \(\finallength{i}/\targetlength{i}\) are in the range \(1 \pm .001\).

\begin{figure*}
   \centering
   \includegraphics{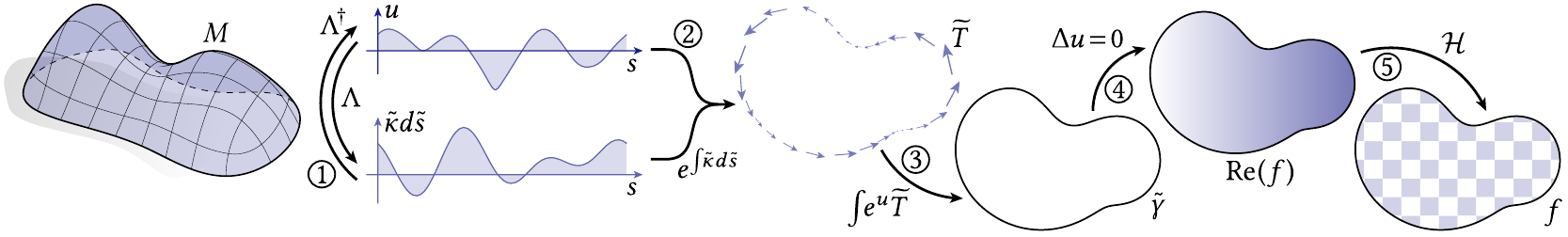}
   \caption{Overview of the basic BFF algorithm.  \(\circled{1}\) Given a surface \(\M\) and either target scale factors \(\u\) or target curvature density \(\tkappa \tds\) along the boundary, the complementary quantity is obtained via the Dirichlet-to-Neumann map \(\Lambda\).  \(\circled{2}\) Curvature density is integrated to obtain unit tangents \(\widetilde{T}\). \(\circled{3}\) Integrating rescaled tangents \(e^\u\widetilde{T}\) yields the target boundary curve \(\tilde{\gamma}\). \(\circled{4}\) The real component of \(\tilde{\gamma}\) is extended harmonically. \(\circled{5}\) The Hilbert transform \(\Htransform\) provides the imaginary coordinate, and hence the final flattening \(\f: \M \to \CC\).~\label{fig:AlgorithmSummary}}
\end{figure*}

\section{Algorithm}
\label{sec:Algorithm}

We now describe the basic boundary first flattening algorithm using tools from \secref{Toolbox}, as summarized in \figref{AlgorithmSummary}.

\bigbreak

\noindent \textbf{Input.} A triangle mesh with disk topology and \emph{either} (i) desired scale factors \(\usf\) \emph{or} (ii) target exterior angles \(\tksf\) at boundary vertices.  In the latter case, angles must sum to \(2\pi\).

\noindent \textbf{Output.} A piecewise linear map \defineDiscreteFlattening \(\fsf: \vertices \to \CC\) which approximates a smooth conformal map with the given boundary data (\(\usf\) or \(\tksf\)).

\noindent \textbf{Algorithm.}
\begin{enumerate}[I.]
   \item \label{stp:ComplementBoundaryData} Compute complementary boundary data (\secref{CherrierFormula}):
      \begin{itemize}
         \item If scale factors \(\usf\) were given, compute compatible angles: \\ \(\phantom{.} \qquad\qquad\qquad \tksf \gets \ksf - \Lambda_\Omega \usf\) \hfill (\secref{DirichletToNeumann}).
         \item If angles \(\tksf\) were given, compute compatible scale factors: \\ \(\phantom{.} \qquad\qquad\qquad \usf \gets \Lambda^\dagger_\Omega (\ksf - \tksf)\) \hfill (\secref{NeumannToDirichlet}).
      \end{itemize}
   \item \label{stp:ReconstructBoundaryCurve} Construct a closed loop \(\tgamma\) that exhibits exterior angles \(\tksf\) and approximates edge lengths \(\targetlength{\ij} := e^{(\usf_i+\usf_j)/2} \edgelength{\ij}\) (\secref{CurveIntegration}).
   \item \label{stp:CurveExtension} Compute the holomorphic extension \(\fsf\) of \(\tgamma\) (\secref{Interpolation}).
\end{enumerate}
\bigbreak
\stepref{ComplementBoundaryData} uses the Cherrier formula to ``explain'' the provided boundary data: if scale factors were given, what must the curvature look like under a conformal flattening (and vice versa)?  The map \(\Lambda_{\Omega}\) (or \smash{\(\Lambda^\dagger_\Omega\)}) is used to evaluate this formula; using a source term \(\Omega\) corresponds to setting the target Gaussian curvature to \smash{\(\widetilde{\Omega} = 0\)}, \ie, \emph{flattening}.  Since the resulting boundary data is compatible with some conformal flattening, the exterior angles \(\smash{\tksf}\) and scaled boundary lengths \(\targetlength{}\) already describe a valid closed loop---modulo a small amount of discretization error which is accounted for by the integration procedure in \stepref{ReconstructBoundaryCurve}. (Here we scale \(\edgelength{}\) by the mean conformal factor \ala\ CETM, though any consistent approximation will work.)  Since length adjustments during integration are very small, the resulting curve \(\tgamma\) remains extremely close to the boundary of some conformal map.  Therefore, in \stepref{CurveExtension} a holomorphic extension of \emph{either} coordinate function \(\re(\tgamma)\) or \(\im(\tgamma)\) will yield an accurate approximation of a conformal map that closely matches the input data---as verified by numerical experiments in \secrefs{Applications}{EvaluationAndComparisons}.

The overall cost is one factorization of a real \(|\vertices| \times |\vertices|\) cotan-Laplace matrix, followed by three backsolves: one to evaluate the map \(\Lambda\) or \(\Lambda^\dagger\) in \stepref{ComplementBoundaryData}; two to compute the holomorphic extension in \stepref{CurveExtension}.  All other operations require only \(O(|\bvertices|)\) work, involving simple closed-form expressions evaluated at boundary vertices.  (See \secref{EvaluationAndComparisons} for detailed performance analysis.)

\section{Applications}
\label{sec:Applications}

We now use the core BFF algorithm to solve several problems in surface parameterization.  The basic question is how to construct appropriate boundary data (scale factors or curvatures) for each task.

\subsection{Automatic Parameterization}
\label{sec:AutomaticParameterization}

In the absence of user-specified criteria, a natural choice of conformal flattening is the one with minimal area distortion.  Springborn \etal~\citeyear[App. E]{Springborn:2008:CET} show that such a map is obtained by prescribing zero scale factors along the boundary (\(\u|_{\bM} = 0\)).  Figures~\ref{fig:teaser}, \ref{fig:AreaDistortion}, and \ref{fig:ModifiedMethodsNiceMesh} show results computed via BFF, which are virtually indistinguishable from those produced by CETM (albeit at far lower cost) and respect features like symmetry even better than SCP (\figref{AreaDistortion}).  These boundary conditions provide the baseline for later comparisons with existing automatic methods (\secref{EvaluationAndComparisons}).

\subsection{Direct Editing}
\label{sec:DirectEditing}

A natural way to edit conformal flattenings with BFF is to directly manipulate the angles \(\theta_{\ij}\) (relative to the real axis) or length \(\targetlength{}\) of target edges \(\ij \in \bedges\).  These values can then be easily converted to curvatures \(\tksf\) and scale factors \(\usf\) per boundary vertex (\resp).  In particular the (integrated) curvatures are simply
\[
   \tksf_i = \theta_{i,i+1} - \theta_{i-1,i}.
\]
Converting target lengths \(\targetlength{\ij}\) (per boundary edge) to target scale factors \(\usf_i\) (per boundary vertex) is more subtle.  One idea is to solve the linear system \(\usf_i + \usf_j = 2\log\targetlength{\ij} - 2\log\edgelength{\ij}\) for scale factors \(\usf\) that satisfy discrete conformal equivalence~\cite{Springborn:2008:CET}.  However, this system does not always have a solution; hence, even metric-based methods like CETM and CMPS cannot (in general) \emph{exactly} prescribe boundary lengths.  We instead use a straightforward numerical approximation: compute scale factors \(\usf_{\ij} := \log(\targetlength{\ij}/\edgelength{\ij})\) per boundary edge, then integrate over dual boundary edges to get
\[
   \usf_j = (\targetlength{\ij} \usf_{\ij} + \targetlength{\jk} \usf_{\jk})/(\targetlength{\ij} + \targetlength{\jk}),
\]
where \(i\), \(j\), and \(k\) are consecutive vertices along the boundary.

\begin{figure}
   \begin{center}
      \includegraphics[width=\columnwidth]{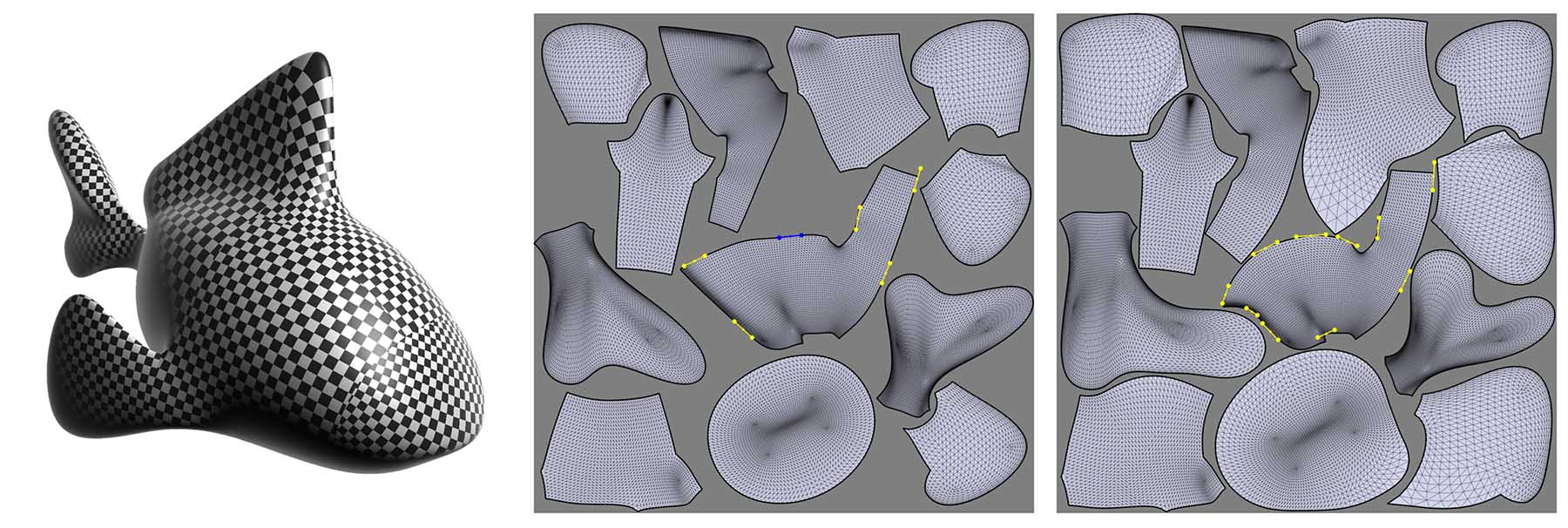}
   \end{center}
   \caption{Unlike conventional tools, our method can be used to interactively and nonrigidly tweak a texture layout while remaining conformal (here aiming for greater use of texture area).~\label{fig:AtlasEditing}}
\end{figure}

As an experiment, we use a spline-based curve editor to nonrigidly pack surface charts into a texture atlas (\figref{AtlasEditing}).  At each vertex, boundary values are sampled from a real-valued Catmull-Rom spline (and normalized to \(2\pi\) in the case of curvatures).  The Dirichlet-to-Neumann map \(\Lambda\) is applied to the current boundary data to switch between angle- and length-based editing.

\subsection{Sharp Corners}
\label{sec:SharpCorners}

\begin{figure}[b]
   \begin{center}
      \includegraphics{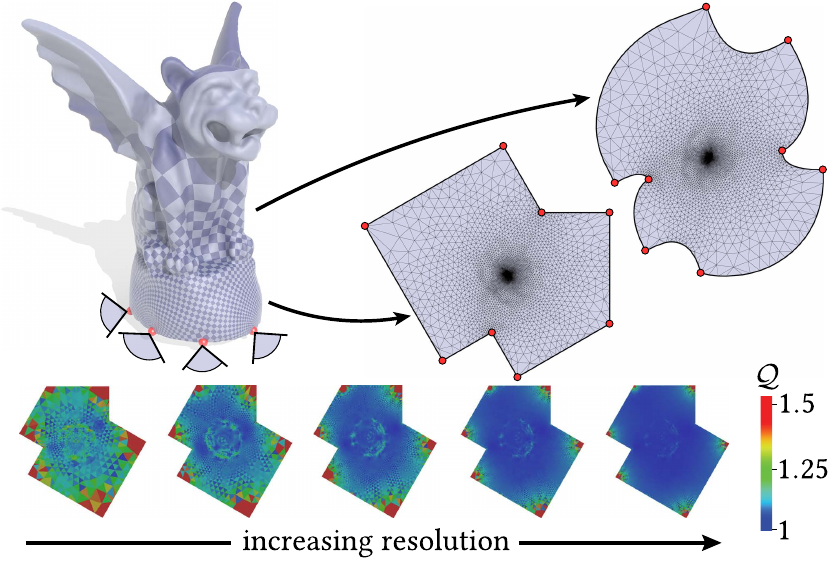}
   \end{center}
   \caption{\figloc{Top left:} Given a collection of points and angles on the boundary, we can map to a region with sharp corners and either straight \figloc{(top center)} or curved \figloc{(top right)} edges. \figloc{Bottom:} This map converges to a perfectly conformal map under refinement.~\label{fig:SharpCornersExample}}
\end{figure}

\setlength{\columnsep}{8pt}%
   \begin{wrapfigure}[14]{l}{50pt}
      \centering
      \includegraphics{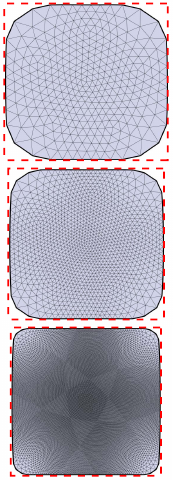}
   \end{wrapfigure}

   For target shapes with sharp corners (like a rectangle) the standard BFF procedure can exhibit undesirable rounding (see inset, \figloc{top}).  Here we can replace the holomorphic extension in \stepref{CurveExtension} with a simple harmonic extension (\secref{Interpolation}), thereby interpolating the polygon \(\tgamma\) reconstructed in \stepref{ReconstructBoundaryCurve}, and \emph{exactly} reproducing the requested angles.  Since \(\tgamma\) approximates the boundary of a conformal map, both procedures still converge to a holomorphic function under refinement (see inset and \figref{SharpCornersExample}, \figloc{bottom}).  Simpler algorithms for mapping to polygons can only handle special cases like rectangles~\cite[Section 4.1]{Zeng:2008:SAC} or are limited to straight edges~\cite{Driscoll:2002:SCM}; compare with the piecewise curved boundary in \figref{SharpCornersExample}, \figloc{top right}.  Note that with piecewise linear maps it is impossible for \emph{any} algorithm to completely eliminate angle distortion near corners, since the initial and target angle sums will differ even under refinement.  

\subsection{Cone Parameterization}
\label{sec:ConeParameterization}

\begin{figure}
   \includegraphics{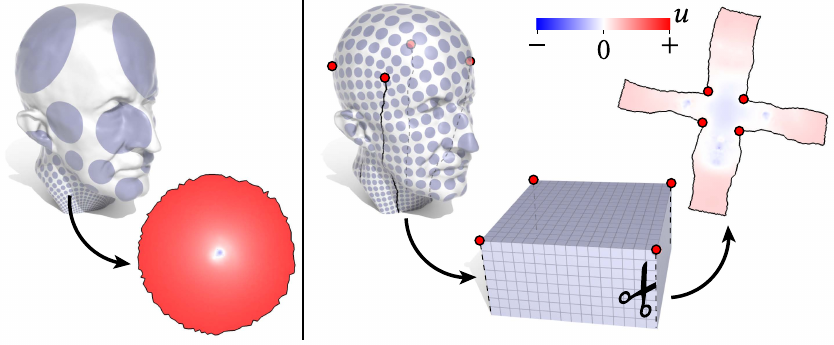}
   \caption{Uneven area distortion (left) can be mitigated by first mapping to a surface that is flat except at a collection of \emph{cone points} (in red), then cutting and unfolding this surface into the plane.  BFF avoids the intermediate metric by directly computing boundary data for the final mapping.\label{fig:ConeParameterization}}
\end{figure}

\begin{figure}
   \includegraphics{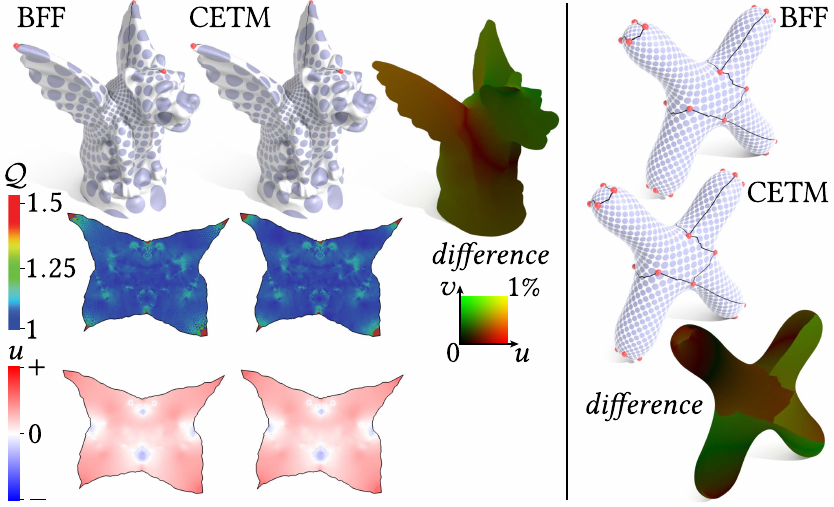}
   \caption{\figloc{Left:} cone parameterizations produced via BFF are virtually identical in quality to those produced by more expensive nonlinear methods like CETM: in addition to similar angle distortion \(\mathcal{Q}\) and scale distortion \(\u\), the map \(\f\) itself differs by less than \(1\%\) relative to the diameter of the image. \figloc{Right:} example with many cone points.\label{fig:ConeComparison}}
\end{figure}

A powerful technique for mitigating area distortion (\figref{ConeParameterization}) is to first map to a \emph{cone surface}, which is flat (\(\K=0\)) away from a collection of isolated \emph{cone points} \cite{Kharevych:2006:DCM}.  After cutting through these points, it can be flattened into the plane without further distortion (\figref{ConeParameterization}, \figloc{right}).  A variety of strategies are available for picking cones~\cite{Kharevych:2006:DCM,BenChen:2008:CFC,Springborn:2008:CET}; we assume they have been specified by the user as target curvatures \defineTargetConeAngles \(\Theta_i\) per interior vertex (mostly zero).  For a closed surface of genus \(g\), \(\Theta\) must sum to \(2\pi(2-2g)\), by Gauss-Bonnet.  To compute a cone flattening of an initial surface \(M_0\), we:
\begin{enumerate}
   \item Solve the Cherrier problem for \(\usf\), with source term \(\Omega - \Theta\).
   \item Cut \(M_0\) into a disk \(\M\) via a cut passing through all cones.
   \item Apply \algref{BoundaryFirstFlattening}, prescribing \(\usf\) values from Step 1 along \(\partial M\).
\end{enumerate}
\citet[Section 3.3]{Springborn:2008:CET} describe one possible cutting procedure.  To obtain a seamless map, we associate only a \emph{single} length degree of freedom \(\finallength{}\) with each pair of corresponding cut edges in \eqref{LengthMatching}; angles are automatically complementary, by Proposition 2, and a harmonic extension (\ala \secref{SharpCorners}) ensures this data is exactly preserved in the final layout.  The resulting maps are nearly identical to those produced by CETM (\figref{ConeComparison}), though we do not exactly preserve length cross ratios.  Unlike CPMS/CETM we never compute (nor need to enforce integrability of) rescaled edge lengths, and can edit cone angles without performing additional factorization; unlike CPMS we guarantee by construction that the map is seamless.  Moreover, since the Laplace matrix is modified only along the cut, fast updates of cone \emph{locations} might also be achieved via low rank updates---see for example Essid \& Solomon~\citeyear{Essid:2017:QRO}.

\subsection{Uniformization}
\label{sec:Uniformization}

\begin{figure}[t]
   \center
   \includegraphics{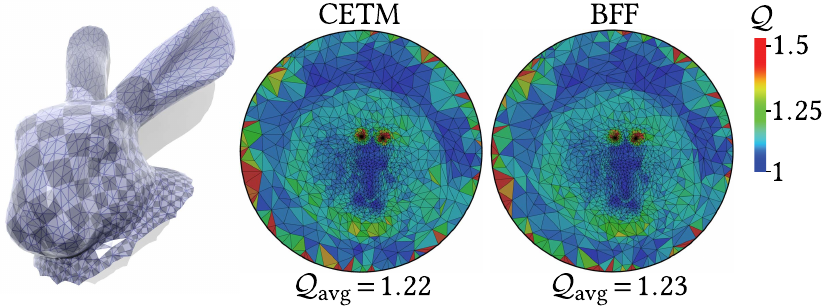}
   \caption{Mapping to the unit disk.  Even on a fairly coarse mesh of 3k triangles \figloc{(left)} we achieve results virtually indistinguishable from nonlinear methods.~\label{fig:Uniformization}}
\end{figure}

A popular use of conformal maps is to provide a common reference domain for comparative data analysis~\cite{Lipman:2011:CWD,Koehl:2015:LFG}.  For surfaces with disk topology, the \emph{uniformization theorem} guarantees the existence of conformal maps to the unit circular disk---any such map has constant boundary curvature \(\kappa = 1\).  However, prescribing a constant curvature \emph{density} yields a shape that is merely convex rather than circular.  Instead, we use a simple fixed-point scheme: if the target curvature \(\tkappa\) were equal to 1, then the target curvature density would be \(\tkappa\tds = \tds = e^\u\ds\), \ie, just the new length density.  Hence, at the \(n\)th iteration we prescribe target angles proportional to the most recent dual edge lengths:
\[
   \tilde{\ksf}^n_i \gets 2\pi \tilde{\ell}^{n-1}_i / \textstyle\sum_{i \in \bvertices} \tilde{\ell}^{n-1}_i.
\]
We then compute a conformal flattening and repeat.  To stabilize this process we average with the previous guess (\ie, \smash{\(\tilde{\ksf}^n \gets \tfrac{1}{2}(\tilde{\ksf}^n + \tilde{\ksf}^{n-1})\)}), using the discrete geodesic curvatures of the input surface as our initial guess.  We also find that a harmonic rather than holomorphic extension (\secref{SharpCorners}) yields better results, especially for domains with jagged boundaries.  In practice we always converge in about 10 iterations---since each iteration involves only backsubstitution the total cost is similar to a single LSCM solve, but produces results nearly identical to nonlinear methods (as depicted in \figref{Uniformization}).  A canonical map can be found by picking the M\"{o}bius transformation that best balances area distortion, \ala\ \cite{Springborn:2005:URP}.

\subsection{Arbitrary Curves}
\label{sec:ArbitraryCurves}

\begin{figure}[b]
   \centering
   \includegraphics{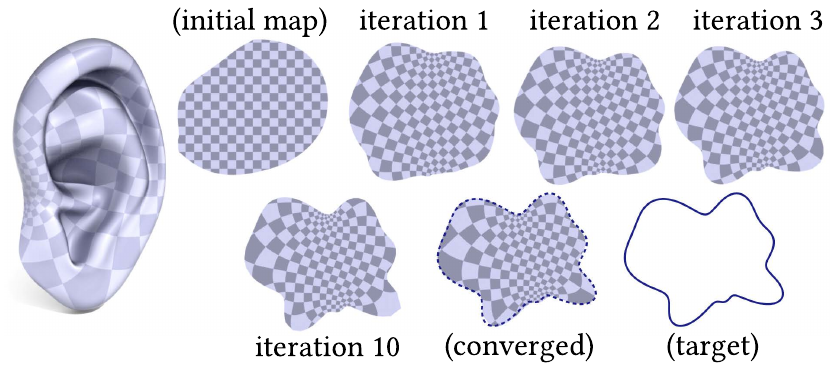}
   \caption{BFF can be used to flatten a surface \figloc{(left)} directly onto a target shape \figloc{(bottom right)} via a simple iterative procedure.  The combined cost of all iterations is not much more than the cost of the initial map.\label{fig:ArbitraryCurve}}
\end{figure}

We can map a surface to an arbitrary target shape using a similar strategy: since a plane curve is determined by its curvature \(\tilde{\kappa}\) (up to rigid motion), we iteratively prescribe the curvature density \(\tilde{\kappa} e^{u^{n-1}} ds\), where \(u^{n-1}\) is our most recent guess for the scale factor.  In the discrete case, let \(\gamma^*(\s): [0,L] \to \CC\) be a desired closed, arc-length parameterized curve; let \(s_i := \sum_{k=1}^i \tilde{\ell}^n_{k,k+1}\) be the cumulative sum of boundary edge lengths at the current iteration, and let \(S := \sum_i s_i\) be the total length.  We first sample \(\gamma^*\) to obtain a polygon with vertices \(\zsf_i := \gamma^*((L/S)s_i)\), \ie, at intervals proportional to our most recent edge lengths.  We then compute the exterior angles of the sampled curve \(\tksf_i = \arg( (z_{i+1}-z_i)/(z_i-z_{i-1}) )\).  Since the sample points \(s_i\) are determined using the most recent length density, these angles provide an estimate for the desired curvature density.  Empirically, this procedure rapidly converges to the target shape (see \figref{ArbitraryCurve}).  Basic as this functionality may seem, it is not provided by any previous conformal flattening algorithm (linear or nonlinear)---the closest comparison is the recent method of Segall and Ben-Chen~\shortcite{Segall:2016:ICC} for in-plane conformal deformations.

\section{Evaluation and Comparisons}
\label{sec:EvaluationAndComparisons}

Here we consider the numerical quality and runtime performance of both linear and nonlinear methods for boundary-controlled conformal flattening, including modifications of existing linear schemes.  To help avoid erroneous comparisons, LSCM, SCP, LinABF, CP, CPMS, and CETM were independently implemented by both authors, and compared with reference implementations wherever possible.  \figref{ModifiedMethodsNiceMesh} confirms that all implementations of boundary-controlled methods produce similar results (albeit at very different costs).

\subsection{Modified Linear Methods}
\label{sec:ModifiedLinearMethods}

Though previous linear conformal flattening methods do not explicitly address boundary control, it is natural to ask whether we simply need to modify their boundary conditions.  Below we explore such modifications for two linear methods (CPMS and LinABF).

\subsection{Boundary-Controlled CPMS}
\label{sec:BoundaryControlledCPMS}

\begin{figure}
   \centering
   \includegraphics{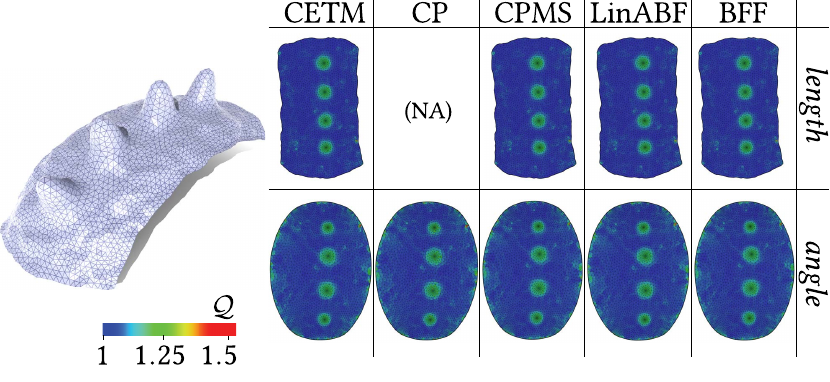}
   \caption{For ideal meshes and smooth boundary data, all five boundary-controlled methods produce near-identical results---albeit with dramatically different update costs.  Here we flatten an optimal Delaunay triangulation using either isometric lengths \figloc{(top)}, or uniform angles \(\tksf_i := \duallength{i}\) \figloc{(bottom)}. Quasi-conformal error \(\Qavg\) was identical (within 0.001) across all methods. \label{fig:ModifiedMethodsNiceMesh}}
\end{figure}

Like BFF, CPMS~\cite{BenChen:2008:CFC} employs a Yamabe-type equation (\secref{CherrierFormula}) to obtain scale information.  There are however several key differences.  First, CPMS does not provide direct control over boundary shape: boundary vertices ``absorb'' curvature via a process that entails a Poisson solve per boundary vertex---even with prefactorization these solves become quite expensive (\citet[Section 4.2]{BenChen:2008:CFC}).  Second, it uses a different layout strategy: whereas BFF need only ensure that boundary data describes a closed loop (\secref{CurveIntegration}), CPMS seeks edge lengths that describe a closed, flat surface.  Rather than satisfy this condition exactly (\ala\ CETM), best-fit vertex positions are found via a least-squares problem~\cite{Sheffer:2004:ABF} which cannot be prefactored since the matrix itself depends on the augmented lengths.  We can of course add boundary control by incorporating \eqref{CherrierFormula}: to control angles we apply the Dirichlet-to-Neumann map (as in \stepref{ComplementBoundaryData} of BFF); to control length we simply set Dirichlet boundary values for \(\u\).  For nice meshes and smooth boundary data this strategy works well (\figref{ModifiedMethodsNiceMesh}) but in general may exhibit artifacts since the least-squares layout does not respect boundary constraints (\figref{ModifiedMethodsAngleControl}).  The amortized cost of editing a map via BFF remains about 30x faster than CPMS with our boundary modification, or about 50x faster than the original method (due to numerous solves at boundary vertices)---see \figref{Timing}.

\subsection{Boundary-Controlled LinABF}
\label{sec:BoundaryControlledLinABF}

\begin{wrapfigure}{r}{53pt}
   \centering
   \includegraphics{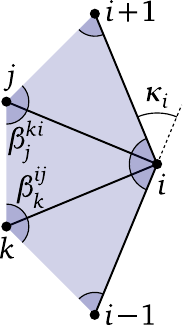}
\end{wrapfigure}
\begin{figure}[t]
   \begin{center}
      \includegraphics{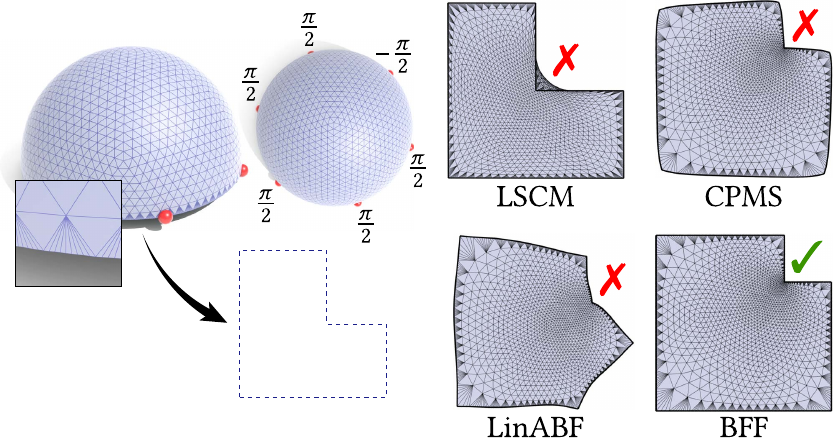}
   \end{center}
   \caption{Hemisphere mapped to an ``L'' shape.  Though existing linear methods can be modified to incorporate angle control, they may exhibit significant artifacts---even on meshes with fairly mild imperfections \figloc{(left)}.  In contrast, exact angle preservation is an algorithmic invariant of BFF.\label{fig:ModifiedMethodsAngleControl}}
\end{figure}

In \emph{angle-based flattening (ABF)} \cite{Sheffer:2001:PFS} a near-flat metric is found by optimizing corner angles \(\tipangle\); a least-squares layout then provides planar vertex positions approximating these angles.  Zayer~\etal~\shortcite{Zayer:2007:LAB} linearize ABF, solving for angle \emph{adjustments} \(\epsilon\) relative to an initial guess \(\tipangle^0\)---this strategy yields results nearly identical to the original algorithm.  We modify this approach in two ways.  First, to prescribe exterior angles \(\tkappa\) we simply add linear constraints \(\sum \tipangle_i^{jk} = \pi - \tkappa_i\) at each boundary vertex \(i\).  Second, to prescribe boundary lengths \(\finallength{\ij}\) (up to global scale) we incorporate the condition
\[
   \prod_{ijk} \sin\tipangle_k^{\ij}/\sin\tipangle_j^{ki} = \finallength{i-1,i}/\finallength{i,i+1}
\]
for each boundary vertex, \ie, we use the law of sines to prescribe the \emph{ratio} of consecutive edge lengths along the boundary (see inset).  Taking the first-order approximation of the logarithm and substituting \(\tipangle_i^0 + \epsilon_i\) for \(\tipangle_i\) in the final system (\ala\ Zayer) then yields linear constraints.  As with CPMS, this strategy works well for nice meshes (\figref{ModifiedMethodsNiceMesh}) but can produce artifacts due to both linearization and the least-squares layout (\figref{ModifiedMethodsAngleControl}).  Moreover, neither the least-squares matrix nor the larger \(3|\faces| \times 3|\faces| \approx 6|\vertices| \times 6|\vertices|\) angle constraint matrix can be prefactored, yielding lower performance than other linear methods (\figref{Timing}).

\begin{figure}[b]
   \centering
   \includegraphics{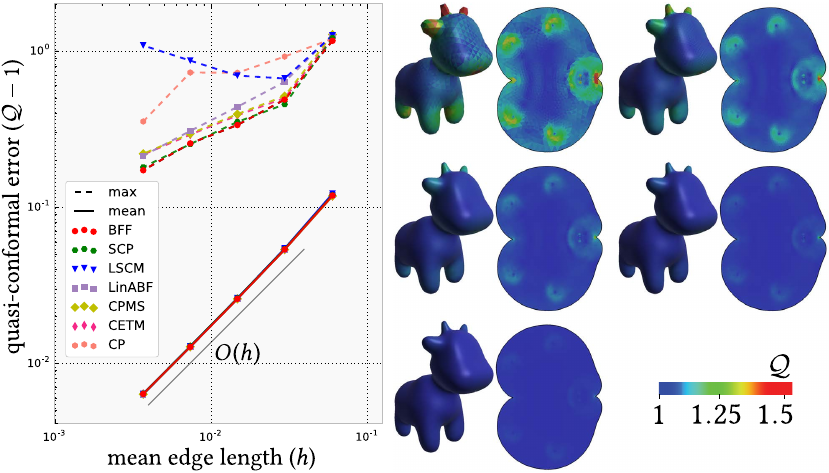}
   \caption{For all methods (both linear and nonlinear) average angle distortion decreases linearly with respect to mean edge length \(h\).  BFF and SCP exhibit the smallest maximum distortion. \figloc{Right:} error distribution for BFF.\label{fig:Convergence}}
\end{figure}

\subsection{Quality and Performance}
\label{sec:QualityAndPerformance}

We measure angle distortion via \emph{quasi-conformal error} \(\Qc\), which is the ratio of singular values of the mapping in each face~\cite{Sander:2001:TMP}; \(\Qc = 1\) is ideal, and \(\Qavg\) denotes the area weighted average over the surface.  Area distortion is measured via the log scale factor \(\u\), which is shifted to have zero mean.

Single-threaded performance was measured on a 3.3GHz Intel Core i7 with 16GB of memory.  All methods were implemented in C++ using double precision.   Linear methods use the supernodal factorization in CHOLMOD~\cite{Chen:2008:ACS}.  To extract the subfactor \(\Lsf_{\II}\) (\secref{PoissonProblems}) we compute a block-preserving reordering via \texttt{cholmod\char`_l\char`_camd}, which has no impact on factorization cost or factor density; overall we achieve a real-world 2x speedup over computing separate Neumann and Dirichlet factors.  For CP we used MOSEK~\cite{Mosek:2010:MOS}.  For CETM we used Newton's method with backtracking line search~\cite[Algorithms 9.2 and 9.5]{Boyd:2004:CO}, which in our tests performed better than L-BFGS or Newton trust region; during editing, using a previous solution as an initial guess did not significantly reduce the number of Newton iterations (especially for large changes to boundary data).

\begin{figure}
   \centering
   \includegraphics{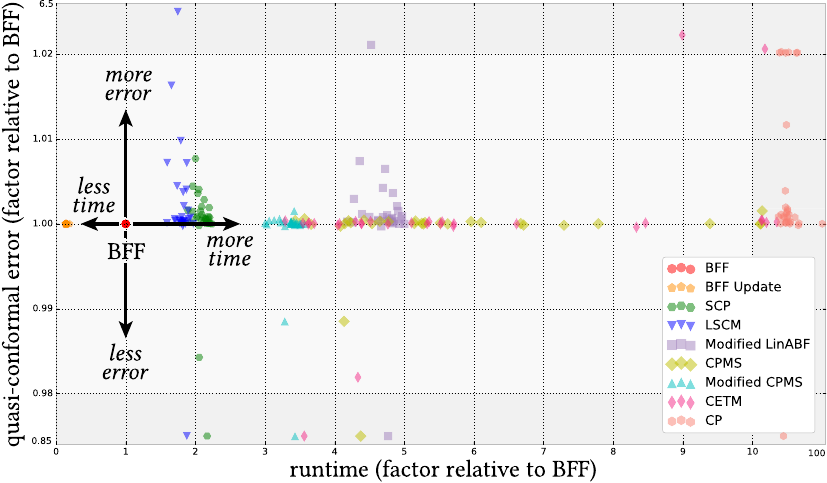}
   \caption{Wall clock time and mean angle distortion relative to BFF, for a large collection of standard meshes (each dot is a single mesh).  Overall, BFF achieves comparable error in less time than all other methods; prefactorization (orange) improves performance by about another order of magnitude.\label{fig:Timing}}
\end{figure}

\figref{Convergence} suggests that our proposed strategy converges to a perfectly conformal map under refinement at the same (linear) rate as all other flattening methods.  \figref{Timing} verifies that real-world timings agree with expected bottlenecks based on the size and number of matrices that must be factored~\cite{Botsch:2005:ELS}.  (Here LSCM/SCP/CPMS/CP use free boundary conditions; all other methods set \(\u|_{\bM} = 0\).)  The takeaway from these experiments is not that any method provides a big win in terms of accuracy (note the narrow vertical range in \figref{Timing}) but simply that BFF does not sacrifice quality for speed.  In particular, by using prefactorization BFF enables conformal flattenings to be edited about 30x quicker than the next-fastest boundary-controlled method (Modified CPMS), and about 50x quicker than the fastest previously-published method (CETM).  This level of improvement provides a qualitative shift in the type of applications that can afford to use sophisticated conformal flattening---\eg, interactive applications, or optimization for computational design~\cite{Konakovic:2016:BDC}.

\subsection{Guarantees and Limitations}
\label{sec:GuaranteesAndLimitations}

\paragraph{Boundary Data.} Two guarantees provided by BFF are (i) exact realization of prescribed angles when using harmonic extension (as discussed in \secref{SharpCorners}) and (ii) exact compatibility of lengths and angles along cuts (as discussed in \secref{ConeParameterization}).  Neither LinABF nor CPMS can provide such guarantees since the least-squares layout step ignores any prescribed boundary data.  Both CP and CETM will exactly satisfy angle constraints so long as optimization converges to an accurate solution.  Exact length constraints are difficult for all methods, as discussed in \secref{DirectEditing}.

\paragraph{Injectivity.} A result of \citet[Theorem 4.1]{Floater:2003:OOP} implies that maps produced by BFF are locally injective as long as (i) prescribed angles \(\ksf\) are positive (describing a convex target), (ii) the discrete Laplacian \(\cotLaplace\) exhibits a maximum principle, and (iii) adjusted edge lengths \(\finallength{}\) are positive.  The Delaunay condition is sufficient (but not necessary) to ensure condition (ii); length adjustments are typically miniscule and hence far from negative (\secref{IntegrationDiscretization}).

\begin{wrapfigure}{r}{140pt}
   \centering
   \includegraphics{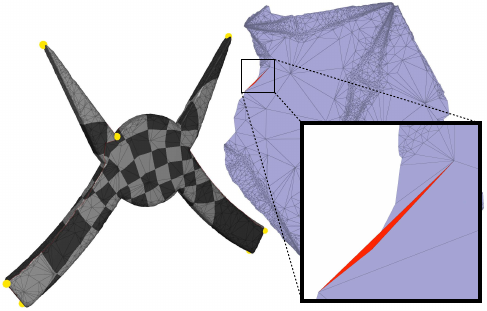}
\end{wrapfigure}
\noindent

In practice we find that flips are quite rare: for instance, 6 of 588 meshes in the SHREC 2011 database had one or two flipped triangles, with a total flipped area on the order of \(10^{-5}\) when normalized to unit radius (see inset); for each mesh we automatically placed eight cones \ala\ \citet{BenChen:2008:CFC}.  Often such flips are easy to fix; one could also perform local Delaunay remeshing, or fall back to a more expensive injective method (see references in~\cite{Smith:2015:BPF}), though such methods may not provide the desired boundary control.

\begin{figure}
   \centering
   \includegraphics[width=\columnwidth]{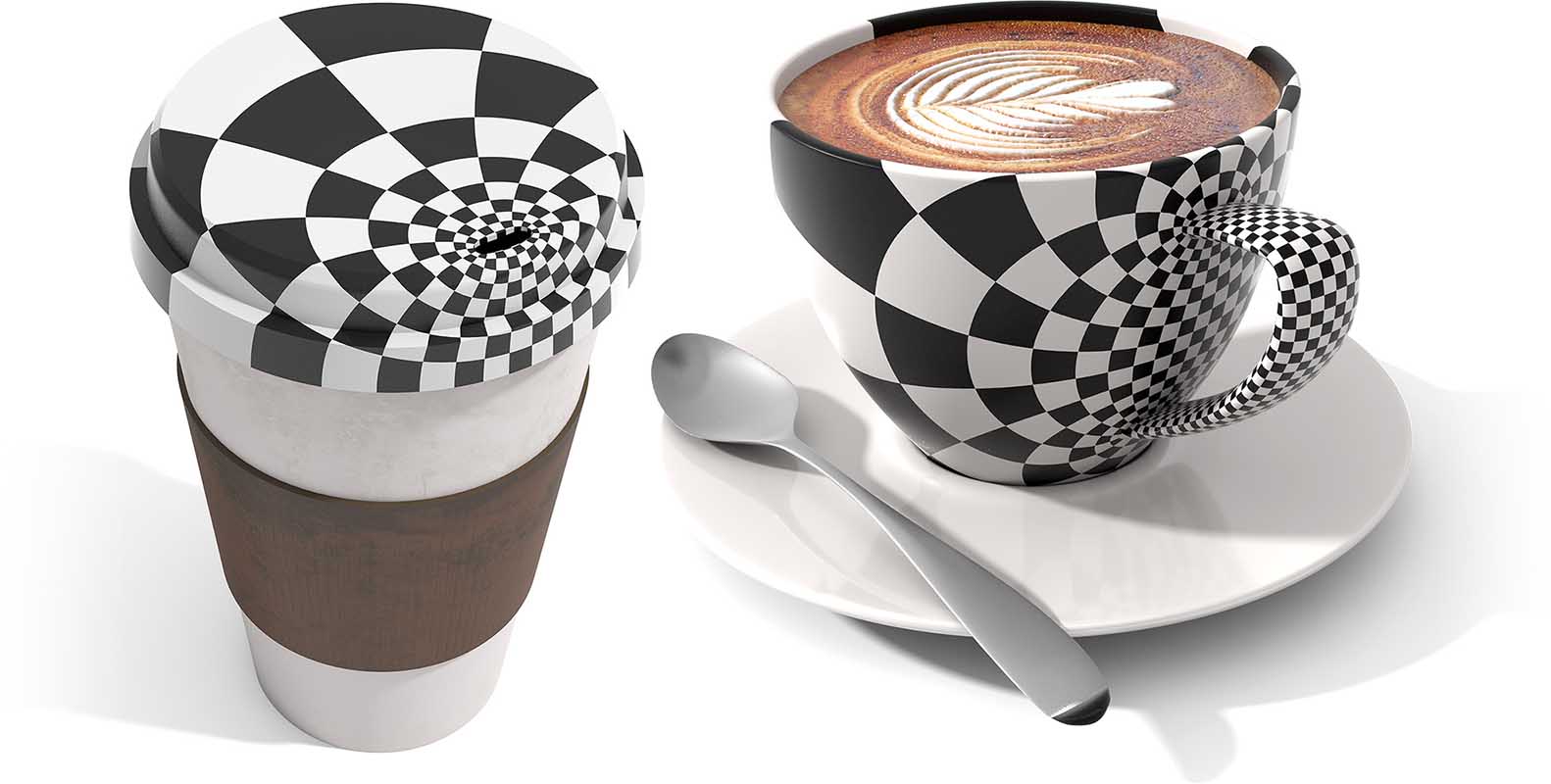}
   \caption{Surfaces with nontrivial topology can be parameterized by cutting them into one or more disks.  Here we cut an annulus \figloc{(left)} and a torus \figloc{(right)} into disks along generators; mapping to rectangles (via prescribed corner angles of \(\pi/2\)) mostly eliminates seams.\label{fig:Coffee}}
\end{figure}

\paragraph{Topology.} For topological spheres one can remove a single vertex star and map to the circular disk, then to the unit sphere via stereographic projection.  For multiply-connected domains like the annulus, the Cherrier formula (\eqref{CherrierFormula}) is still valid but describes only an intrinsic flattening that may not admit a planar layout (consider cutting the tip from a circular cone); moreover, the Hilbert transform (\secref{HilbertTransform}) is valid only for disk-like domains.  In practice, of course, BFF can be used to flatten any surface by cutting it into one or more disks (Figs. \ref{fig:Coffee}, \ref{fig:AtlasEditing}, and \ref{fig:ConeParameterization}).

\paragraph{Target Geometry.} BFF provides flattening only over Euclidean domains (possibly with cone singularities), unlike methods based on discrete Ricci flow~\cite{Jin:2008:DSR,Springborn:2008:CET} which can target spherical or hyperbolic geometry.  However, since the Cherrier formula is valid for any target curvature \(\widetilde{K}\) it would be interesting to generalize the boundary-first strategy---here one might apply an iterative strategy akin to \secref{Uniformization} to prescribe pointwise Gaussian curvature (rather than Gaussian curvature density).

\section{Acknowledgements} Thanks to Tim Davis for help with block reordering and supernodal subfactor extraction in CHOLMOD.  This work was funded in part by a gift from Autodesk, Inc.

% Bibliography
%\bibliographystyle{ACM-Reference-Format}
%\bibliography{BoundaryFirstFlattening}

\begin{thebibliography}{00}

%%% the Web address.

\ifx \showCODEN    \undefined \def \showCODEN     #1{\unskip}     \fi
\ifx \showDOI      \undefined \def \showDOI       #1{{\tt DOI:}\penalty0{#1}\ }
  \fi
\ifx \showISBNx    \undefined \def \showISBNx     #1{\unskip}     \fi
\ifx \showISBNxiii \undefined \def \showISBNxiii  #1{\unskip}     \fi
\ifx \showISSN     \undefined \def \showISSN      #1{\unskip}     \fi
\ifx \showLCCN     \undefined \def \showLCCN      #1{\unskip}     \fi
\ifx \shownote     \undefined \def \shownote      #1{#1}          \fi
\ifx \showarticletitle \undefined \def \showarticletitle #1{#1}   \fi
\ifx \showURL      \undefined \def \showURL       #1{#1}          \fi
% invisible to TeX
\providecommand\bibfield[2]{#2}
\providecommand\bibinfo[2]{#2}
\providecommand\natexlab[1]{#1}
\providecommand\showeprint[2][]{arXiv:#2}

\bibitem[\protect\citeauthoryear{Alexa and Wardetzky}{Alexa and
  Wardetzky}{2011}]%
        {Alexa:2011:DLG}
\bibfield{author}{\bibinfo{person}{Marc Alexa} {and} \bibinfo{person}{Max
  Wardetzky}.} \bibinfo{year}{2011}\natexlab{}.
\newblock
  \showarticletitle{\href{http://doi.acm.org/10.1145/2010324.1964997}{Discrete
  Laplacians on General Polygonal Meshes}}.
\newblock \bibinfo{journal}{{\em ACM Trans.\ Graph.\/}} \bibinfo{volume}{30},
  \bibinfo{number}{4} (\bibinfo{year}{2011}).
\newblock


\bibitem[\protect\citeauthoryear{Alliez, Verdi\`{e}re, Devillers, and
  Isenburg}{Alliez et~al\mbox{.}}{2003}]%
        {Alliez:2003:ISR}
\bibfield{author}{\bibinfo{person}{Pierre Alliez}, \bibinfo{person}{\'{E}ric
  Colin~de Verdi\`{e}re}, \bibinfo{person}{Olivier Devillers}, {and}
  \bibinfo{person}{Martin Isenburg}.} \bibinfo{year}{2003}\natexlab{}.
\newblock
  \showarticletitle{\href{http://dl.acm.org/citation.cfm?id=829510.830318}{Isotropic
  Surface Remeshing}}. In \bibinfo{booktitle}{{\em Proc.\ Shape Modeling
  International}}.
\newblock


\bibitem[\protect\citeauthoryear{ApS}{ApS}{2010}]%
        {Mosek:2010:MOS}
\bibfield{author}{\bibinfo{person}{Mosek ApS}.}
  \bibinfo{year}{2010}\natexlab{}.
\newblock \showarticletitle{The MOSEK optimization software}.
\newblock \bibinfo{journal}{{\em http://www.mosek.com\/}}
  (\bibinfo{year}{2010}).
\newblock


\bibitem[\protect\citeauthoryear{Aubin}{Aubin}{1998}]%
        {Aubin:1998:SNP}
\bibfield{author}{\bibinfo{person}{T. Aubin}.} \bibinfo{year}{1998}\natexlab{}.
\newblock \bibinfo{booktitle}{{\em
  \href{https://books.google.com/books?id=l2nEoSxpHfoC}{Some Nonlinear Problems
  in Riemannian Geometry}}}.
\newblock \bibinfo{publisher}{Springer Berlin}.
\newblock


\bibitem[\protect\citeauthoryear{Ben-Chen, Gotsman, and Bunin}{Ben-Chen
  et~al\mbox{.}}{2008}]%
        {BenChen:2008:CFC}
\bibfield{author}{\bibinfo{person}{Mirela Ben-Chen}, \bibinfo{person}{Craig
  Gotsman}, {and} \bibinfo{person}{Guy Bunin}.}
  \bibinfo{year}{2008}\natexlab{}.
\newblock
  \showarticletitle{\href{https://diglib.eg.org/handle/10.2312/CGF.v27i2pp449-458}{Conformal
  Flattening by Curvature Prescription and Metric Scaling}}.
\newblock \bibinfo{journal}{{\em CG\ Forum\/}} (\bibinfo{year}{2008}).
\newblock


\bibitem[\protect\citeauthoryear{{Bobenko}, {Pinkall}, and
  {Springborn}}{{Bobenko} et~al\mbox{.}}{2010}]%
        {Bobenko:2010:DCM}
\bibfield{author}{\bibinfo{person}{A. {Bobenko}}, \bibinfo{person}{U.
  {Pinkall}}, {and} \bibinfo{person}{B. {Springborn}}.}
  \bibinfo{year}{2010}\natexlab{}.
\newblock
  \showarticletitle{\href{http://adsabs.harvard.edu/abs/2010arXiv1005.2698B}{Discrete
  conformal maps and ideal hyperbolic polyhedra}}.
\newblock \bibinfo{journal}{{\em ArXiv e-prints\/}} (\bibinfo{year}{2010}).
\newblock


\bibitem[\protect\citeauthoryear{{Bobenko} and {G{\"u}nther}}{{Bobenko} and
  {G{\"u}nther}}{2015}]%
        {Bobenko:2015:DCA}
\bibfield{author}{\bibinfo{person}{A.~I. {Bobenko}} {and} \bibinfo{person}{F.
  {G{\"u}nther}}.} \bibinfo{year}{2015}\natexlab{}.
\newblock
  \showarticletitle{\href{http://adsabs.harvard.edu/abs/2015arXiv150505673B}{Discrete
  complex analysis on planar quad-graphs}}.
\newblock \bibinfo{journal}{{\em ArXiv e-prints\/}} (\bibinfo{year}{2015}).
\newblock


\bibitem[\protect\citeauthoryear{Botsch, Bommes, and Kobbelt}{Botsch
  et~al\mbox{.}}{2005}]%
        {Botsch:2005:ELS}
\bibfield{author}{\bibinfo{person}{Mario Botsch}, \bibinfo{person}{David
  Bommes}, {and} \bibinfo{person}{Leif Kobbelt}.}
  \bibinfo{year}{2005}\natexlab{}.
\newblock
  \showarticletitle{\href{http://dx.doi.org/10.1007/11537908_5}{Efficient
  Linear System Solvers for Mesh Processing}}. In \bibinfo{booktitle}{{\em
  Proc.\ IMA\ Int.\ Conf.\ Math.\ Surf.}} \bibinfo{pages}{62--83}.
\newblock


\bibitem[\protect\citeauthoryear{Boyd and Vandenberghe}{Boyd and
  Vandenberghe}{2004}]%
        {Boyd:2004:CO}
\bibfield{author}{\bibinfo{person}{Stephen Boyd} {and} \bibinfo{person}{Lieven
  Vandenberghe}.} \bibinfo{year}{2004}\natexlab{}.
\newblock \bibinfo{booktitle}{{\em Convex Optimization}}.
\newblock \bibinfo{publisher}{Cambridge University Press}.
\newblock
\showISBNx{0521833787}


\bibitem[\protect\citeauthoryear{Bunin}{Bunin}{2008}]%
        {Bunin:2008:CTU}
\bibfield{author}{\bibinfo{person}{Guy Bunin}.}
  \bibinfo{year}{2008}\natexlab{}.
\newblock \showarticletitle{\href{https://arxiv.org/abs/cs/0609078}{A continuum
  theory for unstructured mesh generation in two dimensions}}. In
  \bibinfo{booktitle}{{\em CAGD}}, Vol.~\bibinfo{volume}{25}.
  \bibinfo{pages}{14--40}.
\newblock


\bibitem[\protect\citeauthoryear{Chao, Pinkall, Sanan, and Schr\"{o}der}{Chao
  et~al\mbox{.}}{2010}]%
        {Chao:2010:SGM}
\bibfield{author}{\bibinfo{person}{Isaac Chao}, \bibinfo{person}{Ulrich
  Pinkall}, \bibinfo{person}{Patrick Sanan}, {and} \bibinfo{person}{Peter
  Schr\"{o}der}.} \bibinfo{year}{2010}\natexlab{}.
\newblock \showarticletitle{\href{http://doi.acm.org/10.1145/1778765.1778775}{A
  Simple Geometric Model for Elastic Deformations}}.
\newblock \bibinfo{journal}{{\em ACM Trans.\ Graph.\/}} \bibinfo{volume}{29},
  \bibinfo{number}{4} (\bibinfo{year}{2010}).
\newblock


\bibitem[\protect\citeauthoryear{Chen, Weber, Keren, and Ben-Chen}{Chen
  et~al\mbox{.}}{2013}]%
        {Chen:2013:PSI}
\bibfield{author}{\bibinfo{person}{Renjie Chen}, \bibinfo{person}{Ofir Weber},
  \bibinfo{person}{Daniel Keren}, {and} \bibinfo{person}{Mirela Ben-Chen}.}
  \bibinfo{year}{2013}\natexlab{}.
\newblock
  \showarticletitle{\href{http://doi.acm.org/10.1145/2461912.2461983}{Planar
  Shape Interpolation with Bounded Distortion}}.
\newblock \bibinfo{journal}{{\em ACM Trans.\ Graph.\/}} \bibinfo{volume}{32},
  \bibinfo{number}{4} (\bibinfo{year}{2013}).
\newblock


\bibitem[\protect\citeauthoryear{Chen, Davis, Hager, and Rajamanickam}{Chen
  et~al\mbox{.}}{2008}]%
        {Chen:2008:ACS}
\bibfield{author}{\bibinfo{person}{Yanqing Chen}, \bibinfo{person}{Timothy~A.
  Davis}, \bibinfo{person}{William~W. Hager}, {and}
  \bibinfo{person}{Sivasankaran Rajamanickam}.}
  \bibinfo{year}{2008}\natexlab{}.
\newblock
  \showarticletitle{\href{http://doi.acm.org/10.1145/1391989.1391995}{Algorithm
  887: CHOLMOD, Supernodal Sparse Cholesky Factorization and Update/Downdate}}.
\newblock \bibinfo{journal}{{\em ACM\ Trans.\ Math.\ Softw.\/}}
  \bibinfo{volume}{35}, \bibinfo{number}{3} (\bibinfo{year}{2008}).
\newblock


\bibitem[\protect\citeauthoryear{Cherrier}{Cherrier}{1984}]%
        {Cherrier:1984:PNN}
\bibfield{author}{\bibinfo{person}{Pascal Cherrier}.}
  \bibinfo{year}{1984}\natexlab{}.
\newblock \showarticletitle{Probl\`{e}ms de Neumann non lin\'{e}aires sur les
  vari\'{e}t\'{e}s riemanniennes}.
\newblock \bibinfo{journal}{{\em Journal of Functional Analysis\/}}
  \bibinfo{volume}{57} (\bibinfo{year}{1984}), \bibinfo{pages}{154--206}.
\newblock


\bibitem[\protect\citeauthoryear{Crane}{Crane}{2013}]%
        {Crane:2013:CGP}
\bibfield{author}{\bibinfo{person}{Keenan Crane}.}
  \bibinfo{year}{2013}\natexlab{}.
\newblock {\em
  \bibinfo{title}{\href{http://thesis.library.caltech.edu/7837/}{Conformal
  Geometry Processing}}}.
\newblock \bibinfo{thesistype}{Ph.D. Dissertation}. \bibinfo{school}{Caltech}.
\newblock


\bibitem[\protect\citeauthoryear{Crane, de~Goes, Desbrun, and
  Schr\"{o}der}{Crane et~al\mbox{.}}{2013}]%
        {Crane:2013:DGP}
\bibfield{author}{\bibinfo{person}{Keenan Crane}, \bibinfo{person}{Fernando de
  Goes}, \bibinfo{person}{Mathieu Desbrun}, {and} \bibinfo{person}{Peter
  Schr\"{o}der}.} \bibinfo{year}{2013}\natexlab{}.
\newblock
  \showarticletitle{\href{http://doi.acm.org/10.1145/2504435.2504442}{Digital
  Geometry Processing with Discrete Exterior Calculus}}. In
  \bibinfo{booktitle}{{\em SIGGRAPH Courses}}.
\newblock


\bibitem[\protect\citeauthoryear{Desbrun, Meyer, and Alliez}{Desbrun
  et~al\mbox{.}}{2002}]%
        {Desbrun:2002:IPS}
\bibfield{author}{\bibinfo{person}{Mathieu Desbrun}, \bibinfo{person}{Mark
  Meyer}, {and} \bibinfo{person}{Pierre Alliez}.}
  \bibinfo{year}{2002}\natexlab{}.
\newblock
  \showarticletitle{\href{https://diglib.eg.org/handle/10.2312/8937}{Intrinsic
  Parameterizations of Surface Meshes}}.
\newblock \bibinfo{journal}{{\em CG\ Forum\/}} (\bibinfo{year}{2002}).
\newblock


\bibitem[\protect\citeauthoryear{Driscoll and Trefethen}{Driscoll and
  Trefethen}{2002}]%
        {Driscoll:2002:SCM}
\bibfield{author}{\bibinfo{person}{T.A. Driscoll} {and} \bibinfo{person}{L.N.
  Trefethen}.} \bibinfo{year}{2002}\natexlab{}.
\newblock \bibinfo{booktitle}{{\em
  \href{http://www.cambridge.org/catalogue/catalogue.asp?isbn=9780521807265}{Schwarz-Christoffel
  Mapping}}}.
\newblock \bibinfo{publisher}{Cambridge University Press}.
\newblock


\bibitem[\protect\citeauthoryear{Essid and Solomon}{Essid and Solomon}{2017}]%
        {Essid:2017:QRO}
\bibfield{author}{\bibinfo{person}{Monty Essid} {and} \bibinfo{person}{Justin
  Solomon}.} \bibinfo{year}{2017}\natexlab{}.
\newblock \showarticletitle{Quadratically-Regularized Optimal Transport on
  Graphs}.
\newblock \bibinfo{journal}{{\em (in submission)\/}} (\bibinfo{year}{2017}).
\newblock


\bibitem[\protect\citeauthoryear{Floater}{Floater}{2003}]%
        {Floater:2003:OOP}
\bibfield{author}{\bibinfo{person}{Michael~S. Floater}.}
  \bibinfo{year}{2003}\natexlab{}.
\newblock \showarticletitle{One-to-one piecewise linear mappings over
  triangulations}.
\newblock \bibinfo{journal}{{\it Math. Comp.}}  \bibinfo{volume}{72}
  (\bibinfo{year}{2003}), \bibinfo{pages}{685--696}.
\newblock


\bibitem[\protect\citeauthoryear{Gu and Yau}{Gu and Yau}{2008}]%
        {Gu:2008:CCG}
\bibfield{author}{\bibinfo{person}{X.D. Gu} {and} \bibinfo{person}{S.T. Yau}.}
  \bibinfo{year}{2008}\natexlab{}.
\newblock \bibinfo{booktitle}{{\em
  \href{https://books.google.com/books?id=4FDvAAAAMAAJ}{Computational Conformal
  Geometry}}}.
\newblock \bibinfo{publisher}{International Press}.
\newblock


\bibitem[\protect\citeauthoryear{Hurdal and Stephenson}{Hurdal and
  Stephenson}{2009}]%
        {Hurdal:2009:DCM}
\bibfield{author}{\bibinfo{person}{Monica~K. Hurdal} {and} \bibinfo{person}{Ken
  Stephenson}.} \bibinfo{year}{2009}\natexlab{}.
\newblock
  \showarticletitle{\href{https://www.ncbi.nlm.nih.gov/pubmed/19049882}{Discrete
  conformal methods for cortical brain flattening}}.
\newblock \bibinfo{journal}{{\em NeuroImage\/}} \bibinfo{volume}{45},
  \bibinfo{number}{1} (\bibinfo{year}{2009}), \bibinfo{pages}{86--98}.
\newblock


\bibitem[\protect\citeauthoryear{Hutchinson}{Hutchinson}{1991}]%
        {Hutchinson:1991:CCM}
\bibfield{author}{\bibinfo{person}{John~E. Hutchinson}.}
  \bibinfo{year}{1991}\natexlab{}.
\newblock
  \showarticletitle{\href{http://projecteuclid.org/euclid.pcma/1416323558}{Computing
  conformal maps and minimal surfaces}}. In \bibinfo{booktitle}{{\em
  Theoretical and Numerical Aspects of Geometric Variational Problems}}.
  \bibinfo{pages}{140--161}.
\newblock


\bibitem[\protect\citeauthoryear{Jin, Gu, Luo, and Kim}{Jin
  et~al\mbox{.}}{2008}]%
        {Jin:2008:DSR}
\bibfield{author}{\bibinfo{person}{Miao Jin}, \bibinfo{person}{Xianfeng Gu},
  \bibinfo{person}{Feng Luo}, {and} \bibinfo{person}{Junho Kim}.}
  \bibinfo{year}{2008}\natexlab{}.
\newblock
  \showarticletitle{\href{http://ieeexplore.ieee.org/document/4483509/}{Discrete
  Surface Ricci Flow}}.
\newblock \bibinfo{journal}{{\em IEEE Trans.\ Vis.\ Comp.\ Graph.\/}}
  \bibinfo{volume}{14} (\bibinfo{year}{2008}), \bibinfo{pages}{1030--1043}.
\newblock


\bibitem[\protect\citeauthoryear{Kharevych, Springborn, and
  Schr\"{o}der}{Kharevych et~al\mbox{.}}{2006}]%
        {Kharevych:2006:DCM}
\bibfield{author}{\bibinfo{person}{Liliya Kharevych}, \bibinfo{person}{Boris
  Springborn}, {and} \bibinfo{person}{Peter Schr\"{o}der}.}
  \bibinfo{year}{2006}\natexlab{}.
\newblock
  \showarticletitle{\href{http://doi.acm.org/10.1145/1138450.1138461}{Discrete
  Conformal Mappings via Circle Patterns}}.
\newblock \bibinfo{journal}{{\em ACM Trans.\ Graph.\/}} \bibinfo{volume}{25},
  \bibinfo{number}{2} (\bibinfo{year}{2006}), \bibinfo{pages}{412--438}.
\newblock


\bibitem[\protect\citeauthoryear{Kim, Hanna, Byun, Santangelo, and Hayward}{Kim
  et~al\mbox{.}}{2012}]%
        {Kim:2012:DRB}
\bibfield{author}{\bibinfo{person}{Jungwook Kim}, \bibinfo{person}{James~A.
  Hanna}, \bibinfo{person}{Myunghwan Byun}, \bibinfo{person}{Christian~D.
  Santangelo}, {and} \bibinfo{person}{Ryan~C. Hayward}.}
  \bibinfo{year}{2012}\natexlab{}.
\newblock
  \showarticletitle{\href{http://science.sciencemag.org/content/335/6073/1201}{Designing
  Responsive Buckled Surfaces by Halftone Gel Lithography}}.
\newblock \bibinfo{journal}{{\em Science\/}} \bibinfo{volume}{335},
  \bibinfo{number}{6073} (\bibinfo{year}{2012}), \bibinfo{pages}{1201--1205}.
\newblock


\bibitem[\protect\citeauthoryear{Koehl and Hass}{Koehl and Hass}{2015}]%
        {Koehl:2015:LFG}
\bibfield{author}{\bibinfo{person}{Patrice Koehl} {and} \bibinfo{person}{Joel
  Hass}.} \bibinfo{year}{2015}\natexlab{}.
\newblock
  \showarticletitle{\href{http://rsif.royalsocietypublishing.org/content/12/113/20150795}{Landmark-free
  geometric methods in biological shape analysis}}.
\newblock \bibinfo{journal}{{\em Journal of The Royal Society Interface\/}}
  \bibinfo{volume}{12}, \bibinfo{number}{113} (\bibinfo{year}{2015}).
\newblock


\bibitem[\protect\citeauthoryear{Konakovic, Crane, Deng, Bouaziz, Piker, and
  Pauly}{Konakovic et~al\mbox{.}}{2016}]%
        {Konakovic:2016:BDC}
\bibfield{author}{\bibinfo{person}{Mina Konakovic}, \bibinfo{person}{Keenan
  Crane}, \bibinfo{person}{Bailin Deng}, \bibinfo{person}{Sofien Bouaziz},
  \bibinfo{person}{Daniel Piker}, {and} \bibinfo{person}{Mark Pauly}.}
  \bibinfo{year}{2016}\natexlab{}.
\newblock
  \showarticletitle{\href{http://doi.acm.org/10.1145/2897824.2925944}{Beyond
  Developable: Computational Design and Fabrication with Auxetic Materials}}.
\newblock \bibinfo{journal}{{\em ACM Trans.\ Graph.\/}} \bibinfo{volume}{35},
  \bibinfo{number}{4} (\bibinfo{year}{2016}).
\newblock


\bibitem[\protect\citeauthoryear{L{\'e}vy, Petitjean, Ray, and
  Maillot}{L{\'e}vy et~al\mbox{.}}{2002}]%
        {Levy:2002:LSC}
\bibfield{author}{\bibinfo{person}{Bruno L{\'e}vy}, \bibinfo{person}{Sylvain
  Petitjean}, \bibinfo{person}{Nicolas Ray}, {and} \bibinfo{person}{J{\'e}rome
  Maillot}.} \bibinfo{year}{2002}\natexlab{}.
\newblock
  \showarticletitle{\href{http://doi.acm.org/10.1145/566654.566590}{Least
  Squares Conformal Maps}}.
\newblock \bibinfo{journal}{{\em ACM Trans.\ Graph.\/}} \bibinfo{volume}{21},
  \bibinfo{number}{3} (\bibinfo{year}{2002}).
\newblock


\bibitem[\protect\citeauthoryear{{Lipman} and {Daubechies}}{{Lipman} and
  {Daubechies}}{2011}]%
        {Lipman:2011:CWD}
\bibfield{author}{\bibinfo{person}{Y. {Lipman}} {and} \bibinfo{person}{I.
  {Daubechies}}.} \bibinfo{year}{2011}\natexlab{}.
\newblock
  \showarticletitle{\href{http://adsabs.harvard.edu/abs/2011arXiv1103.4408L}{Conformal
  Wasserstein distances: comparing surfaces in polynomial time}}.
\newblock \bibinfo{journal}{{\em ArXiv e-prints\/}} (\bibinfo{year}{2011}).
\newblock


\bibitem[\protect\citeauthoryear{Lipman, Kim, and Funkhouser}{Lipman
  et~al\mbox{.}}{2012}]%
        {Lipman:2012:SFF}
\bibfield{author}{\bibinfo{person}{Yaron Lipman}, \bibinfo{person}{Vladimir~G.
  Kim}, {and} \bibinfo{person}{Thomas~A. Funkhouser}.}
  \bibinfo{year}{2012}\natexlab{}.
\newblock
  \showarticletitle{\href{http://doi.acm.org/10.1145/2231816.2231822}{Simple
  Formulas for Quasiconformal Plane Deformations}}.
\newblock \bibinfo{journal}{{\em ACM Trans.\ Graph.\/}} \bibinfo{volume}{31},
  \bibinfo{number}{5} (\bibinfo{year}{2012}).
\newblock


\bibitem[\protect\citeauthoryear{Liu, Zhang, Xu, Gotsman, and Gortler}{Liu
  et~al\mbox{.}}{2008}]%
        {Liu:2008:LAM}
\bibfield{author}{\bibinfo{person}{Ligang Liu}, \bibinfo{person}{Lei Zhang},
  \bibinfo{person}{Yin Xu}, \bibinfo{person}{Craig Gotsman}, {and}
  \bibinfo{person}{Steven Gortler}.} \bibinfo{year}{2008}\natexlab{}.
\newblock
  \showarticletitle{\href{http://dl.acm.org/citation.cfm?id=1731309.1731336}{A
  Local/Global Approach to Mesh Parameterization}}. In \bibinfo{booktitle}{{\em
  Proc.\ Symp.\ Geom.\ Proc.}} \bibinfo{pages}{1495--1504}.
\newblock


\bibitem[\protect\citeauthoryear{Luo}{Luo}{2004}]%
        {Luo:2004:CYF}
\bibfield{author}{\bibinfo{person}{Feng Luo}.} \bibinfo{year}{2004}\natexlab{}.
\newblock
  \showarticletitle{\href{http://www.worldscientific.com/doi/abs/10.1142/S0219199704001501}{Combinatorial
  Yamabe Flow on Surfaces}}.
\newblock \bibinfo{journal}{{\em Comm.\ Contemp.\ Math.\\/}}
  \bibinfo{volume}{06}, \bibinfo{number}{05} (\bibinfo{year}{2004}),
  \bibinfo{pages}{765--780}.
\newblock


\bibitem[\protect\citeauthoryear{MacNeal}{MacNeal}{1949}]%
        {MacNeal:1949:SPD}
\bibfield{author}{\bibinfo{person}{Richard MacNeal}.}
  \bibinfo{year}{1949}\natexlab{}.
\newblock {\em
  \bibinfo{title}{\href{http://thesis.library.caltech.edu/1529/}{The Solution
  of Partial Differential Equations by means of Electrical Networks}}}.
\newblock \bibinfo{thesistype}{Ph.D. Dissertation}. \bibinfo{school}{Caltech}.
\newblock


\bibitem[\protect\citeauthoryear{Mercat}{Mercat}{2001}]%
        {Mercat:2001:DRS}
\bibfield{author}{\bibinfo{person}{Christian Mercat}.}
  \bibinfo{year}{2001}\natexlab{}.
\newblock
  \showarticletitle{\href{http://dx.doi.org/10.1007/s002200000348}{Discrete
  Riemann Surfaces and the Ising Model}}.
\newblock \bibinfo{journal}{{\em Comm.\ Math.\ Phys.\/}} \bibinfo{volume}{218},
  \bibinfo{number}{1} (\bibinfo{year}{2001}), \bibinfo{pages}{177--216}.
\newblock


\bibitem[\protect\citeauthoryear{Mullen, Tong, Alliez, and Desbrun}{Mullen
  et~al\mbox{.}}{2008}]%
        {Mullen:2008:SCP}
\bibfield{author}{\bibinfo{person}{Patrick Mullen}, \bibinfo{person}{Yiying
  Tong}, \bibinfo{person}{Pierre Alliez}, {and} \bibinfo{person}{Mathieu
  Desbrun}.} \bibinfo{year}{2008}\natexlab{}.
\newblock
  \showarticletitle{\href{http://dl.acm.org/citation.cfm?id=1731309.1731335}{Spectral
  Conformal Parameterization}}. In \bibinfo{booktitle}{{\em Proc.\ Symp.\
  Geom.\ Proc.}} \bibinfo{pages}{1487--1494}.
\newblock


\bibitem[\protect\citeauthoryear{Myles and Zorin}{Myles and Zorin}{2013}]%
        {Myles:2013:CCG}
\bibfield{author}{\bibinfo{person}{Ashish Myles} {and} \bibinfo{person}{Denis
  Zorin}.} \bibinfo{year}{2013}\natexlab{}.
\newblock
  \showarticletitle{\href{http://doi.acm.org/10.1145/2461912.2461970}{Controlled-distortion
  Constrained Global Parametrization}}.
\newblock \bibinfo{journal}{{\em ACM Trans.\ Graph.\/}} \bibinfo{volume}{32},
  \bibinfo{number}{4} (\bibinfo{date}{July} \bibinfo{year}{2013}).
\newblock


\bibitem[\protect\citeauthoryear{Polthier}{Polthier}{2000}]%
        {Polthier:2000:CHM}
\bibfield{author}{\bibinfo{person}{Konrad Polthier}.}
  \bibinfo{year}{2000}\natexlab{}.
\newblock
  \showarticletitle{\href{https://projecteuclid.org/euclid.em/1062620735}{Conjugate
  Harmonic Maps and Minimal Surfaces}}.
\newblock \bibinfo{journal}{{\em Preprint No. 446, TU-Berlin, SFB 288\/}}
  (\bibinfo{year}{2000}).
\newblock


\bibitem[\protect\citeauthoryear{Sander, Snyder, Gortler, and Hoppe}{Sander
  et~al\mbox{.}}{2001}]%
        {Sander:2001:TMP}
\bibfield{author}{\bibinfo{person}{Pedro~V. Sander}, \bibinfo{person}{John
  Snyder}, \bibinfo{person}{Steven~J. Gortler}, {and} \bibinfo{person}{Hugues
  Hoppe}.} \bibinfo{year}{2001}\natexlab{}.
\newblock
  \showarticletitle{\href{http://doi.acm.org/10.1145/383259.383307}{Texture
  Mapping Progressive Meshes}}. In \bibinfo{booktitle}{{\em Proc.\
  ACM/SIGGRAPH}}. \bibinfo{pages}{409--416}.
\newblock


\bibitem[\protect\citeauthoryear{Sarkar, Yin, Gao, Luo, and Gu}{Sarkar
  et~al\mbox{.}}{2009}]%
        {Rik:2009:GRG}
\bibfield{author}{\bibinfo{person}{Rik Sarkar}, \bibinfo{person}{Xiaotian Yin},
  \bibinfo{person}{Jie Gao}, \bibinfo{person}{Feng Luo}, {and}
  \bibinfo{person}{Xianfeng~David Gu}.} \bibinfo{year}{2009}\natexlab{}.
\newblock
  \showarticletitle{\href{http://doi.acm.org/10.1145/1602165.1602178}{Greedy
  routing with guaranteed delivery using Ricci flows.}}. In
  \bibinfo{booktitle}{{\em IPSN}}. \bibinfo{pages}{121--132}.
\newblock


\bibitem[\protect\citeauthoryear{Segall and Ben-Chen}{Segall and
  Ben-Chen}{2016}]%
        {Segall:2016:ICC}
\bibfield{author}{\bibinfo{person}{Aviv Segall} {and} \bibinfo{person}{Mirela
  Ben-Chen}.} \bibinfo{year}{2016}\natexlab{}.
\newblock
  \showarticletitle{\href{https://diglib.eg.org/handle/10.1111/cgf12961}{Iterative
  Closest Conformal Maps between Planar Domains}}.
\newblock \bibinfo{journal}{{\em CG\ Forum\/}} (\bibinfo{year}{2016}).
\newblock


\bibitem[\protect\citeauthoryear{Segall, Vantzos, and Ben-Chen}{Segall
  et~al\mbox{.}}{2016}]%
        {Segall:2016:HSF}
\bibfield{author}{\bibinfo{person}{Aviv Segall}, \bibinfo{person}{Orestis
  Vantzos}, {and} \bibinfo{person}{Mirela Ben-Chen}.}
  \bibinfo{year}{2016}\natexlab{}.
\newblock
  \showarticletitle{\href{http://dl.acm.org/citation.cfm?id=2982818.2982831}{Hele-Shaw
  Flow Simulation with Interactive Control}}. In \bibinfo{booktitle}{{\em In
  Proc.\ Symp.\ Comp.\ Anim.}} \bibinfo{pages}{85--95}.
\newblock


\bibitem[\protect\citeauthoryear{Sheffer and de~Sturler}{Sheffer and
  de~Sturler}{2001}]%
        {Sheffer:2001:PFS}
\bibfield{author}{\bibinfo{person}{A. Sheffer} {and} \bibinfo{person}{E. de
  Sturler}.} \bibinfo{year}{2001}\natexlab{}.
\newblock
  \showarticletitle{\href{http://dx.doi.org/10.1007/PL00013391}{Parameterization
  of Faceted Surfaces for Meshing using Angle-Based Flattening}}.
\newblock \bibinfo{journal}{{\em Engineering with Computers\/}}
  \bibinfo{volume}{17}, \bibinfo{number}{3} (\bibinfo{year}{2001}),
  \bibinfo{pages}{326--337}.
\newblock


\bibitem[\protect\citeauthoryear{Sheffer, L{\'e}vy, Mogilnitsky, and
  Bogomyakov}{Sheffer et~al\mbox{.}}{2005}]%
        {Sheffer:2004:ABF}
\bibfield{author}{\bibinfo{person}{Alla Sheffer}, \bibinfo{person}{Bruno
  L{\'e}vy}, \bibinfo{person}{Maxim Mogilnitsky}, {and}
  \bibinfo{person}{Alexander Bogomyakov}.} \bibinfo{year}{2005}\natexlab{}.
\newblock
  \showarticletitle{\href{http://doi.acm.org/10.1145/1061347.1061354}{ABF++\,:
  Fast and Robust Angle Based Flattening}}.
\newblock \bibinfo{journal}{{\em ACM Trans.\ Graph.\/}} \bibinfo{volume}{24},
  \bibinfo{number}{2} (\bibinfo{year}{2005}).
\newblock


\bibitem[\protect\citeauthoryear{Sheffer, Praun, and Rose}{Sheffer
  et~al\mbox{.}}{2006}]%
        {Sheffer:2006:MPM}
\bibfield{author}{\bibinfo{person}{Alla Sheffer}, \bibinfo{person}{Emil Praun},
  {and} \bibinfo{person}{Kenneth Rose}.} \bibinfo{year}{2006}\natexlab{}.
\newblock \showarticletitle{\href{http://dx.doi.org/10.1561/0600000011}{Mesh
  Parameterization Methods and Their Applications}}.
\newblock \bibinfo{journal}{{\em Found. Trends. Comput. Graph. Vis.\/}}
  \bibinfo{volume}{2}, \bibinfo{number}{2} (\bibinfo{year}{2006}),
  \bibinfo{pages}{105--171}.
\newblock


\bibitem[\protect\citeauthoryear{Smith and Schaefer}{Smith and
  Schaefer}{2015}]%
        {Smith:2015:BPF}
\bibfield{author}{\bibinfo{person}{Jason Smith} {and} \bibinfo{person}{Scott
  Schaefer}.} \bibinfo{year}{2015}\natexlab{}.
\newblock \showarticletitle{\href{http://doi.acm.org/10.1145/2766947}{Bijective
  Parameterization with Free Boundaries}}.
\newblock \bibinfo{journal}{{\em ACM Trans.\ Graph.\/}} \bibinfo{volume}{34},
  \bibinfo{number}{4} (\bibinfo{date}{July} \bibinfo{year}{2015}).
\newblock


\bibitem[\protect\citeauthoryear{Springborn}{Springborn}{2005}]%
        {Springborn:2005:URP}
\bibfield{author}{\bibinfo{person}{Boris Springborn}.}
  \bibinfo{year}{2005}\natexlab{}.
\newblock \showarticletitle{A Unique Representation of Polyhedral Types.
  Centering via M\"{ö}bius Transformations}.
\newblock \bibinfo{journal}{{\em Math. Z.\/}}  \bibinfo{volume}{249}
  (\bibinfo{year}{2005}), \bibinfo{pages}{513--517}.
\newblock


\bibitem[\protect\citeauthoryear{Springborn, Schr\"{o}der, and
  Pinkall}{Springborn et~al\mbox{.}}{2008}]%
        {Springborn:2008:CET}
\bibfield{author}{\bibinfo{person}{Boris Springborn}, \bibinfo{person}{Peter
  Schr\"{o}der}, {and} \bibinfo{person}{Ulrich Pinkall}.}
  \bibinfo{year}{2008}\natexlab{}.
\newblock
  \showarticletitle{\href{http://doi.acm.org/10.1145/1360612.1360676}{Conformal
  Equivalence of Triangle Meshes}}.
\newblock \bibinfo{journal}{{\em ACM Trans.\ Graph.\/}} \bibinfo{volume}{27},
  \bibinfo{number}{3} (\bibinfo{year}{2008}).
\newblock


\bibitem[\protect\citeauthoryear{Weber and Gotsman}{Weber and Gotsman}{2010}]%
        {Weber:2010:CCM}
\bibfield{author}{\bibinfo{person}{Ofir Weber} {and} \bibinfo{person}{Craig
  Gotsman}.} \bibinfo{year}{2010}\natexlab{}.
\newblock
  \showarticletitle{\href{http://doi.acm.org/10.1145/1778765.1778815}{Controllable
  Conformal Maps for Shape Deformation and Interpolation}}.
\newblock \bibinfo{journal}{{\em ACM Trans.\ Graph.\/}} \bibinfo{volume}{29},
  \bibinfo{number}{4} (\bibinfo{year}{2010}).
\newblock


\bibitem[\protect\citeauthoryear{Weber and Zorin}{Weber and Zorin}{2014}]%
        {Weber:2014:LIP}
\bibfield{author}{\bibinfo{person}{Ofir Weber} {and} \bibinfo{person}{Denis
  Zorin}.} \bibinfo{year}{2014}\natexlab{}.
\newblock \showarticletitle{Locally injective parametrization with arbitrary
  fixed boundaries}.
\newblock \bibinfo{journal}{{\em ACM Trans.\ Comp.\ Sys.\/}}
  \bibinfo{volume}{33}, \bibinfo{number}{4} (\bibinfo{year}{2014}).
\newblock


\bibitem[\protect\citeauthoryear{Zayer, L{\'e}vy, and Seidel}{Zayer
  et~al\mbox{.}}{2007}]%
        {Zayer:2007:LAB}
\bibfield{author}{\bibinfo{person}{Rhaleb Zayer}, \bibinfo{person}{Bruno
  L{\'e}vy}, {and} \bibinfo{person}{Hans-Peter Seidel}.}
  \bibinfo{year}{2007}\natexlab{}.
\newblock
  \showarticletitle{\href{http://dl.acm.org/citation.cfm?id=1281991.1282010}{Linear
  Angle Based Parameterization}}. In \bibinfo{booktitle}{{\em Proc.\ Symp.\
  Geom.\ Proc.}} \bibinfo{pages}{135--141}.
\newblock


\bibitem[\protect\citeauthoryear{Zeng, Lui, Gu, and Yau}{Zeng
  et~al\mbox{.}}{2008}]%
        {Zeng:2008:SAC}
\bibfield{author}{\bibinfo{person}{Wei Zeng}, \bibinfo{person}{Lok~Ming Lui},
  \bibinfo{person}{Xianfeng Gu}, {and} \bibinfo{person}{Shing-Tung Yau}.}
  \bibinfo{year}{2008}\natexlab{}.
\newblock
  \showarticletitle{\href{http://projecteuclid.org/euclid.maa/1254492833}{Shape
  Analysis by Conformal Modules}}.
\newblock \bibinfo{journal}{{\em Meth.\ Appl.\ Anal.\/}} \bibinfo{volume}{15},
  \bibinfo{number}{4} (\bibinfo{year}{2008}), \bibinfo{pages}{539--556}.
\newblock


\bibitem[\protect\citeauthoryear{Zhong, Shuai, Jin, and Guo}{Zhong
  et~al\mbox{.}}{2014}]%
        {Zhong:2014:ASM}
\bibfield{author}{\bibinfo{person}{Zichun Zhong}, \bibinfo{person}{Liang
  Shuai}, \bibinfo{person}{Miao Jin}, {and} \bibinfo{person}{Xiaohu Guo}.}
  \bibinfo{year}{2014}\natexlab{}.
\newblock
  \showarticletitle{\href{http://dx.doi.org/10.1016/j.gmod.2014.03.011}{Anisotropic
  Surface Meshing with Conformal Embedding}}.
\newblock \bibinfo{journal}{{\em Graph. Models\/}} \bibinfo{volume}{76},
  \bibinfo{number}{5} (\bibinfo{year}{2014}), \bibinfo{pages}{468--483}.
\newblock


\end{thebibliography}

\appendix

\section{Exact Angle Sums}
\label{app:ExactAngleSums}

\begin{wrapfigure}{r}{70pt}
   \centering
   \includegraphics{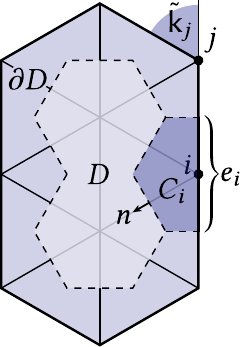}
\end{wrapfigure}

\textsc{Proposition 1.}
\label{prp:ExactAngleSum}
\emph{For any given scale factors \(\usf\), the complementary curvatures \(\tksf\) computed by \stepref{ComplementBoundaryData} of BFF will always sum to exactly \(2\pi\).}

\vspace{-.5\baselineskip}\mbox{}

\noindent\textsc{Proof.} Integrating the Cherrier formula over a boundary dual cell \(\dualcell_i\) yields \( (\cotLaplace\usf)_i + \hsf_i = \Omega_i \) (since \(\widetilde{\Omega} = 0\) for a flattening).  Since \(\Omega_i = 0\) at boundary vertices (\secref{DiscreteCurvature}), the new boundary curvatures can be expressed as
   \[
      \tksf_i = \ksf_i - \hsf_i = \ksf_i + (\cotLaplace\usf)_i = \textstyle\int_{\dbedge_i} \!\!\kappa\ \ds - \textstyle\int_{\partial \dualcell_i \setminus \dbedge_i} \tfrac{\partial \u}{\partial \n}\ \ds,\phantom{mmmm}
   \]
where \(\dbedge_i\) is the dual boundary edge at \(i\). Their sum is then
   \[
      \sum_i \textstyle\int_{e_i} \kappa\ \ds - \sum_i \textstyle\int_{\partial \dualcell_i \setminus \dbedge_i} \tfrac{\partial \u}{\partial \n}\ \ds = \textstyle\int_{\bM} \kappa\ \ds + \textstyle\int_{\partial D} \tfrac{\partial \u}{\partial \n}\ \ds,
   \]
where \(D\) is the union of all interior dual cells (see inset)---note the change in orientation, and cancellation due to equal and opposite normal derivatives in adjacent cells.  Applying Gauss-Bonnet to the first term and the divergence theorem to the second, we get
\[
   2\pi\chi - \textstyle\int_{\M} \K\ \dA - \textstyle\int_D \Delta \u\ \dA,
\]
where for a disk \(\chi = 1\).  But since \(\Delta \u = \K\) (and curvature is nonzero only on \(D\)), we are left with just \(2\pi\). \hfill\(\square\)\bigbreak

\label{app:ComplementaryAngles}

\begin{wrapfigure}[7]{l}{140pt}
   \centering
   \includegraphics{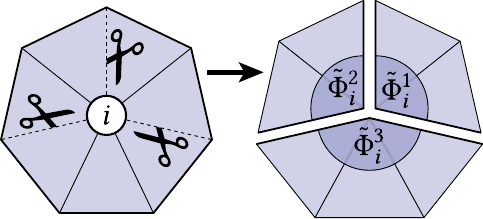}
\end{wrapfigure}

\textsc{Proposition 2.}
\label{prp:ComplementaryAngles}
\emph{Target angles \(\tksf\) computed as in \secref{ConeParameterization} will sum to exactly the desired cone angles \(\Theta\).}

\vspace{-.5\baselineskip}\mbox{}

\noindent\textsc{Proof.} Suppose our cut partitions an interior vertex \(i\) into several regions, and let \(\tilde{\Phi}_i^p\) be the sum of interior angles at \(i\) in the \(p\)th region (see inset).  Recall that \(\tksf = \ksf - h\) (\eqref{DiscreteCurvatureChange}), hence
\[
   \sum_p \tksf_i^p = \sum_p \ksf_i^p - \sum_p \textstyle\int_{\partial C_i^p \setminus e_i^p} \tfrac{\partial \u}{\partial \n}\ \ds
\]
Since \(\tksf\) and \(\tilde{\Phi}\) are supplementary angles, we get
\[
   \sum_p (\pi - \tilde{\Phi}_i^p) = \sum_p (\pi - \Phi_i^p) - \textstyle\int_{C_i} \Delta \u\ \dA,
\]
where \(\partial \dualcell_i\) is the cell boundary on the original (uncut) mesh.  Canceling \(\pi\) terms and recalling that for our cone parameterization \(\Delta u = \Omega - \Theta\) (\secref{ConeParameterization}), we get the desired result:
\[
   \sum_p \tilde{\Phi}_i^p = (2\pi-\Omega_i) - (\Omega_i-\Theta_i) = 2\pi - \Theta_i.
\]
\hfill\(\square\)

%Numerical experiments agree with these observations, up to floating-point error.

\vspace{-\baselineskip}
\section{Pseudocode}
\label{app:Pseudocode}

Pseudocode for the basic BFF algorithm (\secref{Algorithm}) is included below.  Boundary vertices are enumerated by cyclic indices \(1, \ldots, |\bvertices|\) throughout.  In lieu of vertex positions, the geometry of the input mesh is specified purely by its edge lengths \(\edgelength{}\), which is useful when the surface is not embedded.  In practice, of course, it may be easier to compute quantities like angles directly from vertex positions.  For an initial implementation it may be easiest to first separately factor the Dirichlet and Neumann Laplace matrices; extracting a subfactor (\ala\ \eqref{BlockFactorization}) will subsequently improve performance.  

\vspace{\baselineskip}
\begin{algorithm}
   \caption{\Proc{BoundaryFirstFlattening}$(\mesh,\edgelength{},[u|\tksf])$}\label{alg:BoundaryFirstFlattening}
   \begin{algorithmic}[1]
      \InputConditions{A manifold triangle mesh \(\mesh = (\bvertices\!\subseteq\!\vertices, \edges, \faces)\) with disk topology, edge lengths \(\edgelength{}: \edges \to \RR_{>0}\) satisfying the triangle inequality in each face, and \emph{either} scale factors \(\usf: \bvertices \to \RR\) \emph{or} exterior angles \(\tksf: \bvertices \to \RR\) that sum to \(2\pi\).}
      \OutputConditions{A flattening \(\fsf: \vertices \to \CC\).}
      \State \(\tipangle \gets \Proc{InteriorAngles}(\mesh,\edgelength{})\)
      \State \(\Omega, \ksf \gets \Proc{DiscreteCurvatures}(\mesh,\tipangle)\)
      \State \(\cotLaplace \gets \Proc{BuildLaplace}(\mesh,\tipangle)\)
      \State \(\Lsf \gets \Proc{CholeskyFactor}(\cotLaplace)\)
      \If {\(\usf\) was given}
      \State \(\tksf \gets \ksf - \Proc{DirichletToNeumann}(\cotLaplace,\Lsf,-\Omega,\usf)\)
      \Else
      \State \(\usf \gets \Proc{NeumannToDirichlet}(\Lsf,-\Omega,\ksf-\tksf)\)
      \EndIf
      \ForEach {\(\ij \in \bedges\)} \(\targetlength{\ij} \gets e^{(\usf_i+\usf_j)/2}\edgelength{\ij}\)
      \EndFor
      \State \(\tgamma \gets \Proc{BestFitCurve}(\mesh,\edgelength,\targetlength{},\tksf)\)
      \State \Return \(\Proc{ExtendCurve}(\mesh,\Lsf,\tgamma)\)
   \end{algorithmic}
\end{algorithm}

\vspace{\baselineskip}
\begin{algorithm}
   \caption{\Proc{BuildLaplace}$(\mesh,\edgelength{})$}\label{alg:BuildLaplace}
   \begin{algorithmic}[1]
      \InputConditions{A mesh \(\mesh\) with edge lengths \(\edgelength{}\).}
      \OutputConditions{A zero-Neumann Laplace matrix \(\cotLaplace \in \RR^{|\vertices| \times |\vertices|}\).}
      \State \(\cotLaplace \gets 0 \in \RR^{|\vertices| \times |\vertices|}\) \Comment{initialize an empty \emph{sparse} matrix}
      \ForEach {\(pqr \in F\)}
      \ForEach {\(\ijk \in \mathcal{C}(pqr)\)} \Comment{\(\mathcal{C}\): circular shifts}
      \State $\cotLaplace_{\ii}, \cotLaplace_{\jj}\ +\!= \tfrac{1}{2} \cot(\tipangle_k^{\ij})$
      \State $\cotLaplace_{\ij}, \cotLaplace_{\ji}\ -\!= \tfrac{1}{2} \cot(\tipangle_k^{\ij})$
      \EndFor
      \EndFor
      \State \Return $\cotLaplace$
   \end{algorithmic}
\end{algorithm}

\vspace{\baselineskip}
\begin{algorithm}
   \caption{\Proc{DirichletToNeumann}$(\cotLaplace,\Lsf,\phi,\gsf)$}\label{alg:DirichletToNeumann}
   \begin{algorithmic}[1]
      \InputConditions{Zero-Neumann Laplace matrix \(\cotLaplace\) and its factorization \(\Lsf\), source term \(\phi\), and Dirichlet boundary data \(\gsf: \bvertices \to \RR\).}
      \OutputConditions{Neumann data \(\hsf: \bvertices \to \RR\).}
      \State \(\Proc{BackSolve}(\Lsf_{\II},\asf,\phi_{\ivertices}-\cotLaplace_{\IB}\gsf)\)
      \State \Return \(\phi_{\bvertices} - \cotLaplace_{\IB}^\transpose \asf - \cotLaplace_{\BB}\gsf\)
   \end{algorithmic}
\end{algorithm}

\vspace{\baselineskip}
\begin{algorithm}
   \caption{\Proc{NeumannToDirichlet}$(\Lsf,\phi,\hsf)$}\label{alg:NeumannToDirichlet}
   \begin{algorithmic}[1]
      \InputConditions{Cholesky factor \(\Lsf\) of the zero-Neumann Laplace matrix, source term \(\phi: \vertices \to \RR\), and Neumann data \(\hsf: \bvertices \to \RR\).}
      \OutputConditions{Dirichlet data \(\gsf: \bvertices \to \RR\).}
      \State \(\Proc{BackSolve}(\Lsf,\asf,\phi - [0;\hsf])\)
      \State \Return \(\asf_{\bvertices}\)
   \end{algorithmic}
\end{algorithm}

\vspace{\baselineskip}
\begin{algorithm}
   \caption{\Proc{BestFitCurve}$(\mesh,\edgelength{},\targetlength{},\tksf)$}\label{alg:BestFitCurve}
   \begin{algorithmic}[1]
      \InputConditions{A disk \((\mesh,\edgelength{})\), desired edge lengths \(\targetlength{}: \bvertices \to \RR_{>0}\), and target exterior angles \(\tksf: \bvertices \to \RR\) that sum to \(2\pi\)}
      \OutputConditions{Vertex positions \(\tgamma: \bvertices \to \CC\).}
      \State \(\tTsf \gets 0 \in \RR^{2 \times |\bvertices|}\) \Comment{dense matrix of tangents}
      \State \(\varphi_{0,1} \gets 0\) \Comment{direction of first tangent}
      \For{\(i = 1, .., |\bvertices|-1\)} \Comment{walk along boundary}
      \State \(\varphi_{i,i+1} \gets \varphi_{i-1,i} + \tksf_i\) \Comment{accumulate exterior angles}
      \State \(\tTsf_{i,i+1} \gets e^{\imath\varphi_{i,i+1}}\) \Comment{put tangent in column of matrix}
      \EndFor
      \State \(\Nsf \gets 0 \in \RR^{|\bvertices| \times |\bvertices|}\) \Comment{boundary mass matrix}
      \ForEach {\(i \in \bvertices\)} \(\Nsf_{\ii} \gets 1/\duallength{i}\) \Comment{\(\duallength{i}\) are dual lengths}
      \EndFor
      \State \(\finallength{} \gets \targetlength{} - \Nsf^{-1}\tTsf^\transpose(\tTsf\Nsf^{-1}\tTsf^\transpose)^{-1}\tTsf \targetlength{}\) \Comment{adjust lengths to close}
      \State \(\tgamma_1 \gets 0 \in \CC\) \Comment{put first vertex at origin}
      \For{\(i = 2, .., |\bvertices|\)} \Comment{for remaining vertices}
      \State \(\tgamma_i \gets \tgamma_{i-1} + \finallength{i-1,i} \tTsf_{i-1,i}\) \Comment accumulate scaled tangents
      \EndFor
      \State \Return \(\tgamma\)
   \end{algorithmic}
\end{algorithm}

\vspace{\baselineskip}
\begin{algorithm}
   \caption{\Proc{ExtendCurve}$(\mesh,\Lsf,\tgamma)$}\label{alg:ExtendCurve}
   \begin{algorithmic}[1]
      \InputConditions{A mesh \(\mesh\) with disk topology, Cholesky factor \(\Lsf\) of the zero-Neumann Laplace matrix, and a closed loop \(\tgamma: \bvertices \to \CC\).}
      \OutputConditions{A flattening \(\fsf: \vertices \to \CC\).}
      \State \(\Proc{BackSolve}(\Lsf_{\II},\asf,-\cotLaplace_{\IB}\re(\tgamma))\) \Comment{harmonic extension}
      \ForEach {\(i \in \bvertices\)} \(\hsf_i \gets \tfrac{1}{2}(\asf_{i+1}-\asf_{i-1})\) \Comment{Hilbert transform}
      \EndFor
      \State \(\Proc{BackSolve}(\Lsf,\bsf,\hsf)\) \Comment{harmonic conjugation}
      \State \Return \(\asf + \imath\bsf\)
   \end{algorithmic}
\end{algorithm}

\vspace{\baselineskip}
All remaining procedures involve either evaluating simple formulas, or calling standard library routines:
\begin{itemize}
   \item \(\textsc{InteriorAngles}(\mesh,\edgelength{})\) --- computes the angles \(\beta\) at each triangle corner.
   \item \(\textsc{DiscreteCurvature}(\mesh,\tipangle)\) --- computes discrete Gauss and geodesic curvature using formulas from \secref{DiscreteCurvature}.  (Note that \(\Omega_i = 0\) at all boundary vertices \(i \in \bvertices\).)
   \item \(\textsc{CholeskyFactor}(\mathsf{A})\) --- returns sparse Cholesky factor \(\Lsf\).
   \item \(\textsc{BackSolve}(\mathsf{L},\xsf,\bsf)\) --- solves \(\mathsf{A}\mathsf{x} = \mathsf{b}\) using the Cholesky factor of \(\mathsf{A}\).
\end{itemize}

\end{document}